\documentclass[journal]{IEEEtran}
\usepackage{amssymb}
\usepackage{amsbsy}
\usepackage{amsmath}
\usepackage{amsthm} 
\usepackage{setspace}
\usepackage{epsfig}
\usepackage{cite}
\usepackage{graphics}
\usepackage{epstopdf}
\usepackage{mathrsfs}
\usepackage[above]{placeins}
\usepackage{algorithm}
\usepackage{algpseudocode}
\usepackage{multirow}
\usepackage{algorithmicx}
\usepackage{tikz}

\begin{document}
\title{\huge{Lattice Gaussian Sampling by Markov Chain Monte Carlo: Bounded Distance Decoding and Trapdoor Sampling}}

\author{Zheng~Wang,~\IEEEmembership{Member, IEEE,}
        and Cong~Ling,~\IEEEmembership{Member, IEEE}
\thanks{This work was presented in part at the IEEE International Symposium on Information Theory (ISIT), Barcelona, Spain,
July 2016.


Z. Wang is with College of Electronic and Information Engineering, Nanjing University of Aeronautics and Astronautics (NUAA), Nanjing, China; C. Ling is with the Department of Electrical and Electronic Engineering, Imperial
College London, London, SW7 2AZ, United Kingdom (e-mail: z.wang@ieee.org, cling@ieee.org).


}}

\maketitle

\begin{abstract}
Sampling from the lattice Gaussian distribution plays an important role in various research fields. In this paper, the Markov chain Monte Carlo (MCMC)-based sampling technique is advanced in several fronts. Firstly, the spectral gap for the independent Metropolis-Hastings-Klein (MHK) algorithm is derived, which is then extended to Peikert's algorithm and rejection sampling; we show that independent MHK exhibits faster convergence. Then, the performance of bounded distance decoding using MCMC is analyzed, revealing a flexible trade-off between the decoding radius and complexity. MCMC is further applied to trapdoor sampling, again offering a trade-off between security and complexity. Finally, the independent multiple-try Metropolis-Klein (MTMK) algorithm is proposed to enhance the convergence rate. The proposed algorithms allow parallel implementation, which is beneficial for practical applications.
\end{abstract}

\IEEEpeerreviewmaketitle


\IEEEpeerreviewmaketitle


%


\section{Introduction}
\IEEEPARstart{N}{o}wadays, lattice Gaussian sampling has drawn a lot of attention in various research fields. In mathematics, Banaszczyk was the first to apply it to prove the transference theorems for lattices \cite{Banaszczyk}. In coding,
lattice Gaussian distribution was employed to obtain the full shaping gain for lattice coding \cite{Forney_89,Kschischang_Pasupathy}, and to achieve the capacity of the Gaussian channel \cite{LB_13}. It was also used to achieve information-theoretic security in the Gaussian wiretap channel \cite{LLBS_12,7360779} and in the bidirectional relay channel \cite{7058433}, respectively. In cryptography, the lattice Gaussian distribution has become a central tool in the construction of many primitives \cite{MicciancioGaussian,Regevlearning,GentryDissertation}. Specifically, lattice Gaussian sampling lies at the core of signature schemes in the Gentry, Peikert and Vaikuntanathan (GPV) paradigm \cite{Trapdoor}.
Furthermore, lattice Gaussian sampling with a suitable variance allows to solve the closest vector problem (CVP) and the shortest vector problem (SVP) \cite{RegevSolvingtheShortestVectorProblem,RegevSolvingtheClosestVectorProblem}.

However, in sharp contrast to the continuous Gaussian density, it is by no means trivial even to sample from a low-dimensional discrete Gaussian distribution. For some special lattices, there are rather efficient algorithms for Gaussian sampling \cite{LB_13,CB16}.
As the default sampling algorithm for general lattices, Klein's algorithm \cite{Klein} only works when the standard deviation $\sigma = \sqrt{\omega(\text{log}\ n)}\cdot\text{max}_{1\leq i \leq n}\|\mathbf{\widehat{b}}_i\|$ \cite{Trapdoor}, where $\omega(\text{log}\ n)$ is a superlogarithmic function, $n$ denotes the lattice dimension and $\mathbf{\widehat{b}}_i$'s are the Gram-Schmidt vectors of the lattice basis $\mathbf{B}$.
Peikert gave an efficient lattice Gaussian sampler in \cite{Peikert10} for parallel implementation, which however requires larger values of $\sigma$.
On the other hand, the lattice Gaussian sampling algorithm proposed by Aggarwal \textit{et al.} in \cite{RegevSolvingtheShortestVectorProblem,RegevSolvingtheClosestVectorProblem} to solve CVP and SVP has a lower bound $2^n$ on both space and time complexity; it actually obtains samples for small $\sigma$ by combining original samples for $\sigma= \sqrt{\omega(\text{log}\ n)}\cdot\text{max}_{1\leq i \leq n}\|\mathbf{\widehat{b}}_i\|$. Although the algorithm in \cite{KirchnerLattice} provides a trade-off between (exponential) time and space complexity, its complexity is still too high to be practical.
%
%
%
%

In order to sample from a target lattice Gaussian distribution with arbitrary $\sigma>0$, Markov chain Monte Carlo (MCMC) methods were introduced in \cite{ZhengWangTIT15}. In principle, it randomly generates the next Markov state conditioned on the previous one; after the burn-in time, which is normally measured by the \emph{mixing time}, the Markov chain will step into a stationary distribution, when samples from the target distribution can be obtained \cite{mixingtimemarkovchain}. It has been demonstrated that Gibbs sampling, which employs univariate conditional sampling to build the chain, yields an ergodic Markov chain \cite{ZhengWangMCMCLatticeGaussian}. In \cite{ZhengWangTIT15}, we proposed an independent Metropolis-Hastings (MH) algorithm incorporating Klein's algorithm (namely, the independent MHK algorithm) to generate a proposal distribution, which is shown to be uniformly ergodic (converging exponentially fast to the stationary distribution). Meanwhile, the associated convergence rate of the Markov chain is derived, resulting in a tractable estimation of the mixing time. Differently from the algorithms of \cite{RegevSolvingtheShortestVectorProblem,RegevSolvingtheClosestVectorProblem,KirchnerLattice}, the independent MHK sampling algorithm only requires polynomial space.
In this paper, we advance the state of the art of MCMC-based lattice Gaussian sampling in several fronts.

Firstly, we refine the analysis and extend the independent MHK algorithm of \cite{ZhengWangTIT15}. We obtain the spectral gap of the transition matrix and demonstrate uniformly ergodicity. We extend the independent MH algorithm to a version where Peikert's algorithm \cite{Peikert10} is used to generate the proposal distribution.
We then compare these MCMC algorithms with rejection sampling from statistics. By deriving their rates of convergence, we show the advantage of the independent MHK. Rejection sampling achieves the same convergence rate only if its normalizing constant is carefully chosen, which is generally rather difficult.

Secondly, we apply the independent MHK algorithm to bounded distance decoding (BDD). BDD is a variant of the CVP where the input is within a certain distance to the lattice. With a careful selection of the standard deviation $\sigma$ during the sampling process, we improve the result of Klein from $\eta=O(1/n)$ to $\eta=O(\sqrt{\log n}/n)$ in terms of $\eta-$BDD\footnote{In $\eta$-BDD ($\eta<1/2$), we are given a lattice basis $\mathbf{B}$ and a query point $\mathbf{c}$, and we are asked to find a lattice point within distance $\eta\cdot\lambda_1$ from the target, where $\lambda_1$ denotes the first minimum of the lattice.}.
References \cite{Yi-KaiLiu2006,DRSD14} achieved a larger value $\eta=O(\sqrt{\log n/n})$, at the expense of a pre-processing stage where Gaussian samples are taken from the dual lattice with standard deviation $\sigma$ equal to its smoothing parameter. However, sampling at the smoothing parameter is in general a difficult problem with no efficient solutions nowadays. For algorithms of general SVP/CVP such as enumeration and sieving, we refer the readers to the comprehensive survey \cite{HPS11}.

Thirdly, we examine the impact of MCMC to trapdoor sampling in the GPV paradigm. In cryptographic applications, the standard deviation $\sigma$ of the sampler is the main parameter governing the security level. Namely, the smaller $\sigma$, the higher security. This is because for a signature system to be secure, it must be hard for an adversary to find lattice points of length about $\sigma\sqrt{n}$. We show that, at moderate costs of increased complexity, MCMC is able to sample with smaller $\sigma$, thereby increasing the security level relative to Klein's algorithm \cite{Trapdoor} and Peikert's algorithm \cite{Peikert10}.

Finally, to improve the convergence rate of the Markov chain, the independent multiple-try Metropolis-Klein (MTMK) algorithm is proposed, which fully exploits the trial samples generated from the proposal distribution. Uniform ergodicity is demonstrated and the enhanced convergence rate is also given. Since independent MHK is only a special case of independent MTMK, the decoding performance can also be improved due to the usage of trial samples.
The proposed sampling algorithm allows a parallel implementation and is easily adopted to MIMO detection to achieve near-optimal performance. With the development of 5G, the demand for large-scale MIMO systems will increase in the next decade, which has triggered research activities towards low complexity decoding algorithms for large-scale MIMO detection \cite{MarzettaMIMO,LarssonMIMO,RusekMIMO}. Therefore, there has been considerable interest in MCMC sampling for the efficient decoding of MIMO systems \cite{HassibiMCMCnew,McmcDatta,MCMCBehrouz,ChenMCMC,XiaodongWangMultilevel,MCMCHaidongZhu}.





The rest of this paper is organized as follows. Section II introduces the lattice Gaussian distribution and briefly reviews the basics of MCMC. In Section III, we derive the spectral gaps of the Markov chains associated with independent MHK and rejection sampling-based lattice Gaussian sampling, and show their uniform ergodicity as well as convergence rates. An extension to Peikert's algorithm is also given.
Then, the decoding complexity of BDD using independent MHK algorithm is derived in Section IV. Section V addresses trapdoor sampling using MCMC. In Section VI, the independent MTMK algorithm is proposed to further strength the convergence performance. Simulation results for MIMO detection are presented in Section VII. Finally, Section VIII concludes the paper.

\emph{Notation:} Matrices and column vectors are denoted by upper
and lowercase boldface letters, and the transpose, inverse, pseudoinverse
of a matrix $\mathbf{B}$ by $\mathbf{B}^T, \mathbf{B}^{-1},$ and
$\mathbf{B}^{\dag}$, respectively. $\mathbf{I}$ denotes the identity matrix. We use $\mathbf{b}_i$ for the $i$th
column of the matrix $\mathbf{B}$, $\mathbf{\widehat{b}}_i$ for the $i$th
Gram-Schmidt vector of the matrix $\mathbf{B}$, $b_{i,j}$ for the entry in the $i$th row
and $j$th column of the matrix $\mathbf{B}$. A symmetric matrix $\mathbf{B}$ is written
as $\mathbf{B}\succ\mathbf{0}$ if it is positive definite. Similarly, we say $\mathbf{B}_1\succ\mathbf{B}_2$
if $(\mathbf{B}_1-\mathbf{B}_2)\succ\mathbf{0}$. $\lceil x \rfloor$ denotes rounding to
the integer closest to $x$. If $x$ is a complex number, $\lceil x \rfloor$
rounds the real and imaginary parts separately.
In addition, we use the standard \emph{small omega} notation $\omega(\cdot)$, i.e., $|\omega(g(n))|>k\cdot|g(n)|$ for every fixed positive number $k>0$. Finally, in this paper, the
computational complexity is measured by the number of Markov moves.

\newtheorem{my1}{Lemma}
\newtheorem{my2}{Theorem}
\newtheorem{my3}{Definition}
\newtheorem{my4}{Proposition}
\newtheorem{my5}{Remark}
\newtheorem{my6}{Conjection}
\newtheorem{my7}{Corollary}

\begin{figure}[t]
\vspace{-2em}
\begin{center}
\hspace{-1em}\includegraphics[width=4in,height=2.4in]{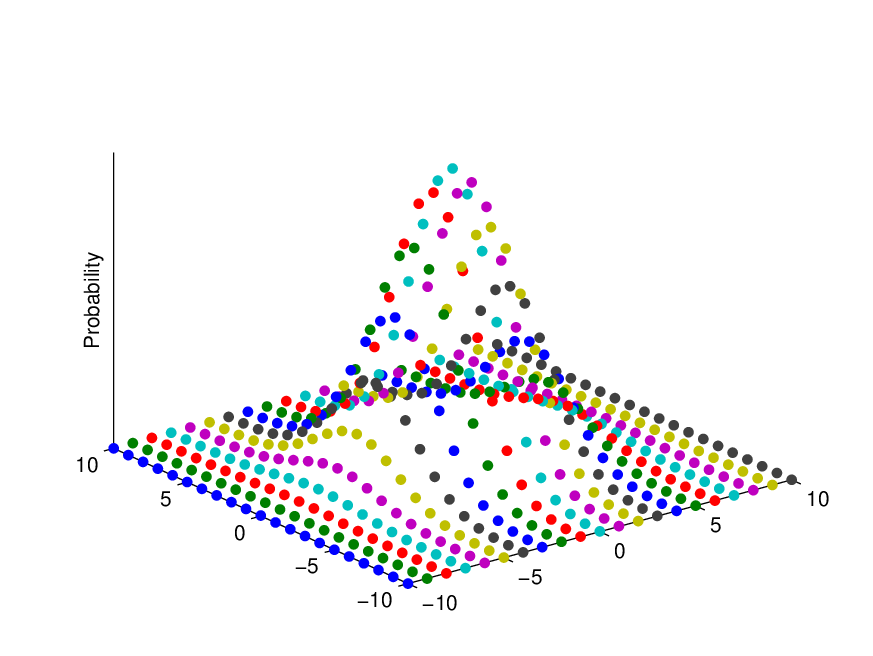}
\end{center}
\vspace{-2em}
  \caption{Illustration of a two-dimensional lattice Gaussian distribution.}
  \label{simulation x}
\end{figure}

\section{Preliminaries}
In this section, we introduce the background and mathematical tools needed
to describe and analyze the proposed lattice Gaussian sampling algorithms.

\subsection{Lattice Gaussian Distribution}
Let matrix $\mathbf{B}=[\mathbf{b}_1,\ldots,\mathbf{b}_n]\subset \mathbb{R}^n$ consist of $n$ linearly independent column vectors. The $n$-dimensional lattice $\Lambda$ generated by $\mathbf{B}$ is defined by
\begin{equation}
\Lambda=\{\mathbf{Bx}: \mathbf{x}\in \mathbb{Z}^n\},
\end{equation}
where $\mathbf{B}$ is called the lattice basis. We define the Gaussian function centered at $\mathbf{c}\in \mathbb{R}^n$ for standard deviation $\sigma>0$ as
\begin{equation}
\rho_{\sigma, \mathbf{c}}(\mathbf{z})=e^{-\frac{\|\mathbf{z}-\mathbf{c}\|^2}{2\sigma^2}},
\end{equation}
for all $\mathbf{z}\in\mathbb{R}^n$. When $\mathbf{c}$ or $\sigma$ are not specified, we assume that they are $\mathbf{0}$ and $1$ respectively. Then, the \emph{discrete Gaussian distribution} over $\Lambda$ is defined as
\begin{equation}
D_{\Lambda,\sigma,\mathbf{c}}(\mathbf{x})=\frac{\rho_{\sigma, \mathbf{c}}(\mathbf{Bx})}{\rho_{\sigma, \mathbf{c}}(\Lambda)}=\frac{e^{-\frac{1}{2\sigma^2}\parallel \mathbf{Bx}-\mathbf{c} \parallel^2}}{\sum_{\mathbf{x} \in \mathbb{Z}^n}e^{-\frac{1}{2\sigma^2}\parallel \mathbf{Bx}-\mathbf{c} \parallel^2}}
\label{lattice gaussian distribution}
\end{equation}
for all $\mathbf{x}\in \mathbb{Z}^n$, where $\rho_{\sigma, \mathbf{c}}(\Lambda)\triangleq \sum_{\mathbf{\mathbf{Bx}}\in\Lambda}\rho_{\sigma, \mathbf{c}}(\mathbf{Bx})$ is just a scaling to obtain a probability distribution.
We remark that this definition differs slightly from the one in \cite{MicciancioGaussian}, where $\sigma$ is
scaled by a constant factor $\sqrt{2\pi}$ (i.e., $s =\sqrt{2\pi}\sigma$). In fact, the discrete Gaussian resembles a continuous Gaussian distribution, but is only defined over a lattice. It has been shown that discrete and continuous Gaussian distributions share similar properties, if the \emph{flatness factor} is small \cite{LLBS_12}.

\subsection{Decoding by Sampling}
Consider the decoding of an $n \times n$ real-valued system. The
extension to the complex-valued system is straightforward
\cite{CongRandom}. Let $\mathbf{x}\in \mathbb{Z}^n$ denote the transmitted signal.
The corresponding received signal $\mathbf{c}$ is given by
\begin{equation}
\mathbf{c}=\mathbf{B}\mathbf{x}+\mathbf{w}
\label{eqn:System Model}
\end{equation}
where $\mathbf{w}$ is the noise vector with zero
mean and variance $\sigma_w^{2}$, $\mathbf{B}$ is an $n\times n$ full column-rank matrix of
channel coefficients. Typically, the conventional maximum likelihood (ML) reads
\begin{equation}
\widehat{\mathbf{x}}=\underset{\mathbf{x}\in \mathbb{Z}^{n}}{\operatorname{arg~min}} \, \|\mathbf{c}-\mathbf{B}\mathbf{x}\|^2
\label{eqn:ML Decoding}
\end{equation}
where $\| \cdot \|$ denotes the Euclidean norm. Clearly, ML decoding corresponds to the CVP. If the received signal $\mathbf{c}$ is the origin, then ML decoding reduces to SVP.

Intuitively, the CVP given in (\ref{eqn:ML Decoding}) can be solved by the lattice Gaussian sampling. Since the distribution is centered at the query point $\mathbf{c}$, the closest lattice point $\mathbf{Bx}$ to $\mathbf{c}$ is assigned the largest sampling probability. Therefore, by multiple samplings, the solution of CVP is the most likely to be returned.
It has been demonstrated that lattice Gaussian sampling is equivalent to CVP via a polynomial-time dimension-preserving reduction \cite{DGStoCVPSVP}.
Meanwhile, by adjusting the sample size, the sampling decoder enjoys a flexible trade-off between performance and complexity.

In \cite{Klein}, Klein introduced an algorithm which performs sampling from a Gaussian-like distribution (see Algorithm 1). It is shown in \cite{Klein,CongRandom,DerandomizedJ} that Klein's algorithm is able to find the closest lattice point when it is close to the input vector: this technique is known as BDD in coding literature, which corresponds to a restricted variant of CVP.

\renewcommand{\algorithmicrequire}{\textbf{Input:}}  
\renewcommand{\algorithmicensure}{\textbf{Output:}} 

\begin{algorithm}[t]
\caption{Klein's Algorithm}
\begin{algorithmic}[1]
\Require
$\mathbf{B}, \sigma, \mathbf{c}$
\Ensure
$\mathbf{Bx}\in\Lambda$
\State let $\mathbf{B}=\mathbf{QR}$ and $\mathbf{c'}=\mathbf{Q}^{\dag}\mathbf{c}$
\For {$i=n$,\ \ldots,\ 1}
\State let $\sigma_i=\frac{\sigma}{|r_{i,i}|}$ and $\widetilde{x}_i=\frac{c'_i-\sum^n_{j=i+1}r_{i,j}x_j}{r_{i,i}}$
\State sample $x_i$ from $D_{\mathbb{Z},\sigma_i,\widetilde{x}_i}$
\EndFor
\State return $\mathbf{Bx}$
\end{algorithmic}
\end{algorithm}

\subsection{Classical MH Algorithms}
In \cite{Hastings1970}, the original Metropolis algorithm was extended to a more general scheme known as the Metropolis-Hastings (MH) algorithm. In particular, let us consider a target invariant distribution $\pi$ together with a proposal distribution
$q(\mathbf{x},\mathbf{y})$. Given the current state $\mathbf{x}$ for Markov chain $\mathbf{X}_t$, a state candidate $\mathbf{y}$ for the next Markov move $\mathbf{X}_{t+1}$ is generated from the proposal distribution $q(\mathbf{x},\mathbf{y})$. Then the acceptance ratio $\alpha$ is computed by
\begin{equation}
\alpha=\text{min}\left\{1,\frac{\pi(\mathbf{y})q(\mathbf{y},\mathbf{x})}{\pi(\mathbf{x})q(\mathbf{x},\mathbf{y})}\right\},
\label{quantity compute}
\end{equation}
and $\mathbf{y}$ will be accepted as the new state with
probability $\alpha$. Otherwise, $\mathbf{x}$ will be retained. In this way, a Markov chain $\{\mathbf{X}_0, \mathbf{X}_1, \ldots\}$ is established with the transition probability $P(\mathbf{x},\mathbf{y})$ as follows:
\begin{equation}
P(\mathbf{x},\mathbf{y})=\begin{cases}q(\mathbf{x},\mathbf{y})\alpha \ \ \ \ \ \ \ \ \ \ \ \ \ \ \text{if}\ \mathbf{y}\neq\mathbf{x}, \\
       1-\sum_{\mathbf{z}\neq\mathbf{x}}q(\mathbf{x},\mathbf{z})\alpha\ \ \text{if}\ \mathbf{y}=\mathbf{x}.
       \end{cases}
\label{eqn:RandomDiscrete}
\end{equation}

It is interesting that in MH algorithms, the proposal distribution $q(\mathbf{x},\mathbf{y})$ can be any fixed distribution from which we can conveniently draw samples. Therefore, there is large freedom in the choice of $q(\mathbf{x},\mathbf{y})$ but it is challenging to find a suitable one with satisfactory convergence. In fact, Gibbs sampling can be viewed as a special case of the MH algorithm, whose proposal distribution is a univariate conditional distribution.

As an important parameter to measure the time required by a
Markov chain to get close to its stationary distribution, the \emph{mixing time} is defined as \cite{mixingtimemarkovchain}
\begin{equation}
t_{\text{mix}}(\epsilon)=\text{min}\{t:\text{max}\|P^t(\mathbf{x}, \cdot)-\pi(\cdot)\|_{TV}\leq \epsilon\},
\label{mixing time}
\end{equation}
where $P^t(\mathbf{x}, \cdot)$ denotes a row of the transition matrix $\mathbf{P}$ for $t$ Markov moves and
$\|\cdot\|_{TV}$ represents the total variation distance.

\subsection{Independent MHK Algorithm}
From the MCMC perspective, lattice Gaussian distribution can be viewed as a complex target distribution lacking direct sampling methods. In order to obtain samples from $D_{\Lambda,\sigma,\mathbf{c}}(\mathbf{x})$, the independent MHK sampling was proposed in \cite{ZhengWangTIT15}. Specifically, a state candidate $\mathbf{y}$ for the next Markov move $\mathbf{X}_{t+1}$ is generated by Klein's algorithm, via the following backward one-dimensional conditional sampling (for $i=n, n-1, \ldots, 1$):
{\allowdisplaybreaks\begin{flalign}
P(y_i|\overline{\mathbf{y}}_{[-i]})&=P(y_i|y_{i+1}, \ldots, y_n)\notag\\
&=\frac{e^{-\frac{1}{2\sigma^2}\parallel \mathbf{\overline{c}'}- \mathbf{\overline{R}\overline{y}} \parallel^2}}{\sum_{y_i \in \mathbb{Z}}e^{-\frac{1}{2\sigma^2}\parallel \mathbf{\overline{c}'} - \mathbf{\overline{R}\overline{y}} \parallel^2}}\notag\\
&=\frac{e^{-\frac{1}{2\sigma^2}\parallel c'_i-\sum_{j=i}^nr_{i,j}y_j \parallel^2}}{\sum_{y_i \in \mathbb{Z}}e^{-\frac{1}{2\sigma^2}\parallel c'_i-\sum_{j=i}^nr_{i,j}y_j \parallel^2}}\notag\\
&=\frac{e^{-\frac{1}{2\sigma_i^2}\parallel y_i-\widetilde{y}_i \parallel^2}}{\sum_{y_i \in \mathbb{Z}}e^{-\frac{1}{2\sigma_i^2}\parallel y_i-\widetilde{y}_i \parallel^2}}\notag\\
&=D_{\mathbb{Z},\sigma_i,\widetilde{y}_i}(y_i),
\label{s5}
\end{flalign}}where $\widetilde{y}_i=\frac{c'_i-\sum^n_{j=i+1}r_{i,j}y_j}{r_{i,i}}$, $\sigma_i=\frac{\sigma}{|r_{i,i}|}$, $\mathbf{c}'=\mathbf{Q}^{\dag}\mathbf{c}$ and $\mathbf{B}=\mathbf{QR}$ by QR decomposition with $\|\mathbf{\widehat{b}}_i\|=|r_{i,i}|$.
Note that $\overline{\mathbf{y}}_{[-i]}=[y_{i+1}, \ldots, y_n]$, $\mathbf{\overline{R}}$, $\mathbf{\overline{c}'}$ and $\mathbf{\overline{y}}$ are the $(n-i+1)$ segments of $\mathbf{R}$, $\mathbf{c}'$ and $\mathbf{y}$ respectively (i.e., $\mathbf{\overline{R}}$ is a $(n-i+1)\times(n-i+1)$ submatrix of $\mathbf{R}$ with $r_{i,i}$ to $r_{n,n}$ in the diagonal).

Given the current state $\mathbf{x}$, the proposal distribution $q(\mathbf{x},\mathbf{y})$ in the independent MHK sampling is given by
{\allowdisplaybreaks\begin{flalign}
q(\mathbf{x},\mathbf{y})&=\prod^n_{i=1} P(y_{n+1-i}|\overline{\mathbf{y}}_{[-(n+1-i)]})\notag\\
&=\frac{\rho_{\sigma, \mathbf{c}}(\mathbf{By})}{\prod^n_{i=1}\rho_{\sigma_{n+1-i},\widetilde{y}_{n+1-i}}(\mathbb{Z})}\notag\\
&=q(\mathbf{y}),
\label{MH proposal density}
\end{flalign}}where the proposal distribution $q(\mathbf{x},\mathbf{y})$ is actually independent of $\mathbf{x}$. Therefore, the connection between two consecutive Markov moves is only due to the decision stage.

With the state candidate $\mathbf{y}$, the acceptance ratio $\alpha$ is obtained by substituting (\ref{MH proposal density}) into (\ref{quantity compute})
\begin{flalign}
\alpha&=\text{min}\left\{1,\frac{\prod^n_{i=1}\rho_{\sigma_{n+1-i}, \widetilde{y}_{n+1-i}}(\mathbb{Z})}{\prod^n_{i=1}\rho_{\sigma_{n+1-i}, \widetilde{x}_{n+1-i}}(\mathbb{Z})}\right\},
\label{MH quantity compute}
\end{flalign}
where $\widetilde{x}_i=\frac{c'_i-\sum^n_{j=i+1}r_{i,j}x_j}{r_{i,i}}$ and we note that $\pi=D_{\Lambda,\sigma,\mathbf{c}}$ in (\ref{quantity compute}) (these notations will be followed throughput the context). The sampling procedure is summarized in Algorithm 2. Note that the initial state $\mathbf{x}_0$ for $\mathbf{X}_{0}$ can be chosen from $\mathbb{Z}^n$ arbitrarily or from
the output of a suboptimal algorithm.

Thanks to the celebrated coupling technique, the uniformly ergodicity was demonstrated in \cite{ZhengWangTIT15}. Nevertheless, the spectral gap of the transition matrix, which serves as an important metric for the mixing time of the underlying Markov chain, has not been determined yet.

%
%
%
%

\begin{algorithm}[t]
\caption{Independent MHK Sampling Algorithm}
\begin{algorithmic}[1]
\Require
$\mathbf{B}, \sigma, \mathbf{c}, \mathbf{x}_0, t_{\text{mix}}(\epsilon)$;
\Ensure $\mathbf{x} \thicksim D_{\Lambda,\sigma,\mathbf{c}}$;
\State let $\mathbf{X}_0=\mathbf{x}_0$
\For {$t=$1,2,\ \ldots,\ }
\State let $\mathbf{x}$ denote the state of $\mathbf{X}_{t-1}$
\State sample $\mathbf{y}$ from the proposal distribution $q(\mathbf{x},\mathbf{y})$ in (\ref{MH proposal density})
\State calculate the acceptance ratio $\alpha(\mathbf{x},\mathbf{y})$ in (\ref{MH quantity compute})
\State generate a sample $u$ from the uniform density $U[0,1]$
\If {$u\leq \alpha(\mathbf{x},\mathbf{y})$}
\State let $\mathbf{X}_t=\mathbf{y}$
\Else
\State $\mathbf{X}_t=\mathbf{x}$
\EndIf
\If {$t\geq t_{\text{mix}}(\epsilon)$ }
\State output $\mathbf{x}$
\EndIf


\EndFor
\end{algorithmic}
\end{algorithm}

\section{Convergence Analysis}
In this section, the spectrum of the Markov chain induced by independent MHK sampling is analyzed, followed by the extensions to Peikert's algorithm and rejection sampling. As a common way to evaluate the mixing time, the \emph{spectral gap} $\gamma=1-|\tau_1|>0$ of the transition matrix is preferred for convergence analysis in MCMC \cite{mixingtimemarkovchain}. Here, $\tau_1$ represents the second largest eigenvalue in magnitude of the transition matrix $\mathbf{P}$ \cite{RapidlyMixingRandall}.

\subsection{Spectral Gap of Independent MHK Algorithm}



\begin{my2}
Given the invariant lattice Gaussian distribution $D_{\Lambda,\sigma,\mathbf{c}}$, the Markov chain induced by independent MHK sampling exhibits a spectral gap
\begin{equation}
\gamma\geq \delta\triangleq\frac{\rho_{\sigma, \mathbf{c}}(\mathbf{\Lambda})}{\prod^n_{i=1}\rho_{\sigma_i}(\mathbb{Z})}.
\label{xxxxabc}
\end{equation}
\end{my2}

\begin{proof}
From (\ref{MH proposal density}) and (\ref{MH quantity compute}), the transition probability $P(\mathbf{x},\mathbf{y})$ of each Markov move in the independent MHK sampling is given by
{\allowdisplaybreaks\begin{flalign}
P(\mathbf{x},\mathbf{y})&=\begin{cases}\text{min}\left\{q(\mathbf{y}),\frac{\pi(\mathbf{y})q(\mathbf{x})}{\pi(\mathbf{x})}\right\} \ \ \ \ \ \ \ \ \ \ \ \ \hspace{.2em}\text{if}\ \mathbf{y}\neq\mathbf{x}, \\
1-\underset{\mathbf{z}\neq\mathbf{x}}{\sum}\text{min}\left\{q(\mathbf{z}),\frac{\pi(\mathbf{z})q(\mathbf{x})}{\pi(\mathbf{x})}\right\}    \ \ \ \ \text{if}\ \mathbf{y}=\mathbf{x}.
\end{cases}\hspace{-1.5em}
\label{b2}
\end{flalign}}

For notational simplicity, we define the \emph{importance weight} $w(\mathbf{x})$ as
\begin{equation}
w(\mathbf{x})=\frac{\pi(\mathbf{x})}{q(\mathbf{x})}.
\label{importance weight}
\end{equation}
Then the transition probability can be rewritten as
{\allowdisplaybreaks\begin{flalign}
P(\mathbf{x},\hspace{-.1em}\mathbf{y})\hspace{-.2em}&=\begin{cases}q(\mathbf{y})\cdot\text{min}\left\{1,\frac{w(\mathbf{y})}{w(\mathbf{x})}\right\} \ \ \ \ \ \ \ \ \ \ \ \ \ \hspace{.2em}\ \text{if}\ \mathbf{y}\neq\mathbf{x}, \\
q(\mathbf{x})\hspace{-.2em}+\hspace{-.4em}\underset{\mathbf{z}\neq\mathbf{x}}{\sum}\hspace{-.1em}q(\mathbf{z})\hspace{-.1em}\cdot\text{max}\hspace{-.1em}\left\{\hspace{-.1em}0, \hspace{-.1em}1\hspace{-.1em}-\hspace{-.1em}\frac{w(\mathbf{z})}{w(\mathbf{x})}\hspace{-.2em}\right\}       \hspace{-.1em}\ \text{if}\ \mathbf{y}=\mathbf{x}.
       \end{cases}\hspace{-1.5em}
\label{eq:b3}
\end{flalign}}

Without loss of generality, we label the countably infinite state space $\Omega=\mathbb{Z}^n$ as $\Omega=\{\mathbf{x}_1,\mathbf{x}_2, \ldots,\mathbf{x}_{\infty}\}$ and assume that these states are sorted according to their importance weights, namely,
\begin{equation}
w(\mathbf{x}_1)\geq w(\mathbf{x}_2)\geq\cdots\geq w(\mathbf{x}_{\infty}).
\label{b8}
\end{equation}


From (\ref{eq:b3}) and (\ref{b8}), the transition matrix $\mathbf{P}$ of the Markov chain can be exactly expressed as
\begin{equation*}
\mathbf{P}\hspace{-0.1em}=\hspace{-0.4em}\left[\begin{array}{ccccc}
                                q(\mathbf{x}_1)+\eta_1 & \frac{\pi(\mathbf{x}_2)}{w(\mathbf{x}_1)} & \frac{\pi(\mathbf{x}_3)}{w(\mathbf{x}_1)} & \cdots & \frac{\pi(\mathbf{x}_{\infty})}{w(\mathbf{x}_1)} \\
                                   q(\mathbf{x}_1)    & q(\mathbf{x}_2)+\eta_2 & \frac{\pi(\mathbf{x}_3)}{w(\mathbf{x}_2)} & \cdots & \frac{\pi(\mathbf{x}_{\infty})}{w(\mathbf{x}_2)}\\
                                 q(\mathbf{x}_1)    &    q(\mathbf{x}_2)    &  q(\mathbf{x}_3)+\eta_3 &\cdots &\frac{\pi(\mathbf{x}_{\infty})}{w(\mathbf{x}_3)}\\
                                 \vdots & \vdots  & \vdots  & \ddots & \vdots\\
                                   q(\mathbf{x}_1)    &    q(\mathbf{x}_2)  &    q(\mathbf{x}_3)  & \cdots & q(\mathbf{x}_{\infty})
                                \end{array}\right]
\end{equation*}
where
\begin{equation}
\eta_j=\sum_{i=j}^{\infty}\left(q(\mathbf{x}_i)-\frac{\pi(\mathbf{x}_i)}{w(\mathbf{x}_j)}\right)
\label{b4}
\end{equation}
stands for the probability of being rejected in the decision stage with the current state $\mathbf{x}_j$ for $\mathbf{X}_t$.

Let $\mathbf{q}=[q(\mathbf{x}_1), q(\mathbf{x}_2), \ldots]^T$ denote the vector of proposal probabilities. Then by decomposition, it follows that
\begin{equation}
\mathbf{P}=\mathbf{G}+\mathbf{e}\mathbf{q}^T,
\label{b5}
\end{equation}
where $\mathbf{e}=[1,1,\ldots]^T$ and $\mathbf{G}$ is an upper triangular matrix of the form
\begin{equation*}
\mathbf{G}\hspace{-0.1em}=\hspace{-0.4em}\left[\begin{array}{cccc}
                                \eta_1 & \frac{\pi(\mathbf{x}_2)}{w(\mathbf{x}_1)}-q(\mathbf{x}_2) & \cdots & \frac{\pi(\mathbf{x}_{\infty})}{w(\mathbf{x}_1)}-q(\mathbf{x}_{\infty}) \\
                                   0    & \eta_2 & \cdots & \frac{\pi(\mathbf{x}_{\infty})}{w(\mathbf{x}_2)}-q(\mathbf{x}_{\infty})\\
                                  \vdots& \vdots  & \ddots &\vdots\\
                                     0  &   0   &  \cdots & 0
                                \end{array}\right].
\end{equation*}

It is well-known that for a Markov chain, the largest eigenvalue of the transition matrix $\mathbf{P}$ always equals 1. Here, as $\mathbf{e}$ is a common right eigenvector for both $\mathbf{P}$ and $\mathbf{P}-\mathbf{G}$, it naturally corresponds to the largest eigenvalue 1. Meanwhile, since the rank of $\mathbf{P}-\mathbf{G}$ is 1, the other eigenvalues of $\mathbf{G}$ are exactly the same as those of $\mathbf{P}$.

Thanks to the ascending order in (\ref{b8}), it is easy to verify that the spectral radius $\tau_1$ is exactly given by
\begin{equation}
\tau_1=\eta_1
\label{xm}
\end{equation}
and
\begin{equation}
1>|\eta_1|\geq|\eta_2|\geq\cdots>0,
\end{equation}
thereby raising the interest of identifying the value of $\tau_1$.

Therefore, according to (\ref{b4}) and (\ref{xm}), we can easily get that
{\allowdisplaybreaks\begin{flalign}
\tau_1&=\sum_{i=1}^{\infty}\left(q(\mathbf{x}_i)-\frac{\pi(\mathbf{x}_i)}{w(\mathbf{x}_1)}\right)\notag\\
&=\sum_{i=1}^{\infty}q(\mathbf{x}_i)-\frac{1}{w(\mathbf{x}_1)}\cdot\sum_{i=1}^{\infty}\pi(\mathbf{x}_i)\notag\\
&=1-\frac{1}{w(\mathbf{x}_1)}\notag\\
&=1-\frac{q(\mathbf{x}_1)}{\pi(\mathbf{x}_1)}.
\label{b6}
\end{flalign}}
In other words, the spectral gap $1-\tau_1$ is exactly captured by the ratio $q(\mathbf{x}_1)/\pi(\mathbf{x}_1)$. Next, we invoke the following Lemma to lower bound the ratio $q(\mathbf{x})/\pi(\mathbf{x})$ for $\mathbf{x}\in\mathbb{Z}^n$.
\begin{my1}[\hspace{-0.005em}\cite{ZhengWangTIT15}]
In the independent MHK algorithm
\begin{equation}
\frac{q(\mathbf{x})}{\pi(\mathbf{x})}\geq \delta
\label{xxxx}
\end{equation}
for all $\mathbf{x}\in \mathbb{Z}^n$, where $\delta$ is defined in \eqref{xxxxabc}.
\end{my1}
The proof is completed by combining \eqref{b6} and \eqref{xxxx}.
\end{proof}

By using the coupling technique, it is shown in \cite{ZhengWangTIT15} that the Markov chain converges exponentially fast to the stationary distribution in total variational distance:
\begin{equation}
\|P^t(\mathbf{x}, \cdot)-D_{\Lambda,\sigma,\mathbf{c}}(\cdot)\|_{TV}\leq (1-\delta)^t,
\label{b65}
\end{equation}
The mixing time of the Markov chain is given by
\begin{equation}
t_{\text{mix}}(\epsilon)=\frac{\text{ln}\hspace{.1em}\epsilon}{\text{ln}(1-\delta)}<(-\text{ln}\hspace{.1em}\epsilon)\cdot\left(\frac{1}{\delta}\right),\ \ \epsilon < 1
\label{upperboundmixing}
\end{equation}
which is proportional to $1/\delta$, and becomes $O(1)$ if $\delta \to 1$.


\subsection{Extension to Peikert's Algorithm}
Klein's sampling algorithm is a randomized variant of Babai's nearest-plane algorithm for lattice decoding \cite{Babai}.
Babai also proposed a simpler decoding scheme by direct rounding\footnote{In communications, Babai's nearest-plane algorithm is known as successive interference cancelation (SIC) while the direct rounding algorithm is referred to as zero-forcing (ZF).}, which was further randomized by Peikert in \cite{Peikert10}. Although Peikert's algorithm requires a higher value of $\sigma$, it is parallelizable and can be more attractive in practical implementation.
In fact, Peikert's algorithm can also be incorporated into the Metropolis-Hastings algorithm to overcome the limitation of $\sigma$.

Specifically, given the standard deviation $\sigma>0$ and a basis $\mathbf{B}$, one chooses a positive definite matrix $\Sigma_1=r^2\cdot\mathbf{B}\mathbf{B}^T\prec\Sigma=\sigma^2\cdot\mathbf{I}$ for $r>0$ (i.e., $\Sigma_2=\Sigma-\Sigma_1$ is positive definite).
Then, the proposed sample $\mathbf{z}\in \Lambda$ is taken from the distribution $\mathbf{c}+\mathbf{z}'+D_{\Lambda-\mathbf{c}-\mathbf{z}', \sqrt{\Sigma_1}}$, where $\mathbf{z}'\in\mathbb{R}^n$ is sampled from the continuous distribution $D_{\sqrt{\Sigma_2}}$. Note the lattice Gaussian distribution $D_{\Lambda-\mathbf{c}-\mathbf{z}', \sqrt{\Sigma_1}}$ is expressed as
\begin{equation}
D_{\Lambda-\mathbf{c}-\mathbf{z}', \sqrt{\Sigma_1}}(\mathbf{Bx})=\frac{\rho_{\sqrt{\Sigma_1}}(\mathbf{Bx}-\mathbf{c}-\mathbf{z}')}{\rho_{\sqrt{\Sigma_1}}(\Lambda-\mathbf{c}-\mathbf{z}')}
\end{equation}
with
\begin{equation}
\rho_{\sqrt{\Sigma_1}}(\mathbf{y})=e^{-\frac{1}{2}\mathbf{y}^T\Sigma_1^{-1}\mathbf{y}}, \quad \mathbf{y}\in\mathbb{R}^n.
\end{equation}
The joint probability distribution of $\mathbf{z}\in \Lambda$ and $\mathbf{z}'\in\mathbb{R}^n$ is given by
\begin{eqnarray}
P(\mathbf{z},\mathbf{z}')\hspace{-.8em}&=&\hspace{-.8em}D_{\Lambda-\mathbf{c}-\mathbf{z}', \sqrt{\Sigma_1}}(\mathbf{z}-\mathbf{c}-\mathbf{z}')\cdot D_{\sqrt{\Sigma_2}}(\mathbf{z}')\notag\\
&=&\frac{\rho_{\sqrt{\Sigma_1}}(\mathbf{z}-\mathbf{c}-\mathbf{z}')}{\rho_{\sqrt{\Sigma_1}}(\Lambda-\mathbf{c}-\mathbf{z}')}\cdot\frac{\rho_{\sqrt{\Sigma_2}}(\mathbf{z}')}{\sqrt{\det(2\pi\Sigma_2)}}\notag\\
&\overset{(a)}{=}&\frac{\rho_{\sqrt{\Sigma_1}}(\mathbf{z}'-\mathbf{z}+\mathbf{c})}{\rho_{\sqrt{\Sigma_1}}(\Lambda-\mathbf{c}-\mathbf{z}')}\cdot\frac{\rho_{\sqrt{\Sigma_2}}(\mathbf{z}')}{\sqrt{\det(2\pi\Sigma_2)}}\notag\\
&\overset{(b)}{=}&\frac{\rho_{\sqrt{\Sigma}}(\mathbf{z}-\mathbf{c})\cdot\rho_{\sqrt{\Sigma_3}}(\mathbf{z}'-\mathbf{c}')}{\rho_{\sqrt{\Sigma_1}}(\Lambda-\mathbf{c}-\mathbf{z}')\cdot\sqrt{\det(2\pi\Sigma_2)}},
\label{Simple rounding proposal density}
\end{eqnarray}
where (a) is due to the symmetry of $\rho_{\sqrt{\Sigma_1}}$, and (b) follows from \cite[Fact 2.1]{Peikert10} with positive definite matrix
$\Sigma_3^{-1}=\Sigma_1^{-1}+\Sigma_2^{-1}$ and $\mathbf{c}'=\Sigma_3\Sigma_1^{-1}(\mathbf{z}-\mathbf{c})$. Consequently, the marginal distribution of $\mathbf{z}$ is
\begin{equation}
P(\mathbf{z})=\frac{\rho_{\sqrt{\Sigma}}(\mathbf{z}-\mathbf{c})}{\sqrt{\det(2\pi\Sigma_2)}}\cdot\int\frac{\rho_{\sqrt{\Sigma_3}}(\mathbf{z}'-\mathbf{c}')}{\rho_{\sqrt{\Sigma_1}}(\Lambda-\mathbf{c}-\mathbf{z}')}d\mathbf{z}'.
\end{equation}
As $\mathbf{z}=\mathbf{Bx}$ for $\mathbf{x}\in\mathbb{Z}^n$, we have
\begin{equation}
P(\mathbf{x})=\frac{\rho_{\sigma,\mathbf{c}}(\mathbf{B}\mathbf{x})}{\sqrt{\det(2\pi\Sigma_2)}}\cdot\int\frac{\rho_{\sqrt{\Sigma_3}}(\mathbf{z}'-\mathbf{c}')}{\rho_{\sqrt{\Sigma_1}}(\Lambda-\mathbf{c}-\mathbf{z}')}d\mathbf{z}'.
\label{quantity compute11a}
\end{equation}

Clearly, $P(\mathbf{\cdot})$ can be used as a proposal distribution $q(\cdot)$ in the MH algorithm to obtain the state candidate $\mathbf{y}\in\mathbb{Z}^n$. In this case,
the acceptance ratio $\alpha$ can be calculated by
\begin{equation}
\alpha=\text{min}\left\{1,\frac{\pi(\mathbf{y})P(\mathbf{x})}{\pi(\mathbf{x})P(\mathbf{y})}\right\},
\label{quantity compute11}
\end{equation}
followed by a decision to accept $\mathbf{X}_{t+1}=\mathbf{y}$ or not. To summarize, its operation procedure is shown in Algorithm 3.

\begin{my1}
In the independent MH algorithm using Peikert's algorithm, there exists a constant $\delta'>0$ such that
\begin{equation}
\frac{q(\mathbf{x})}{\pi(\mathbf{x})}\geq \delta'
\label{xxxx12348}
\end{equation}
for all $\mathbf{x}\in \mathbb{Z}^n$, where
\begin{equation}
\delta'=\frac{\rho_{\sigma,\mathbf{c}}(\Lambda)}{\rho_{r}(\mathbb{Z}^n)}\cdot\frac{r^n}{\sigma^n}\cdot{|\det(\mathbf{B})|}.
\label{xxxxabcd}
\end{equation}

\end{my1}

\begin{proof}
To start with, we have
{\allowdisplaybreaks\begin{flalign}
\frac{q(\mathbf{x})}{\pi(\mathbf{x})}&=\frac{\rho_{\sigma,\mathbf{c}}(\mathbf{B}\mathbf{x})}{\sqrt{\det(2\pi\Sigma_2)}}\cdot\int\frac{\rho_{\sqrt{\Sigma_3}}(\mathbf{z}'-\mathbf{c}')}{\rho_{\sqrt{\Sigma_1}}(\Lambda-\mathbf{c}-\mathbf{z}')}d\mathbf{z}'\cdot\frac{\rho_{\sigma,\mathbf{c}}(\Lambda)}{\rho_{\sigma,\mathbf{c}}(\mathbf{B}\mathbf{x})}\notag\\
&\overset{(c)}{\geq}\frac{\rho_{\sigma,\mathbf{c}}(\Lambda)}{\sqrt{\det(2\pi\Sigma_2)}}\cdot\frac{1}{\rho_{\sqrt{\Sigma_1}}(\Lambda)}\cdot\int\rho_{\sqrt{\Sigma_3}}(\mathbf{z}'-\mathbf{c}')d\mathbf{z}'\notag\\
&=\frac{\rho_{\sigma,\mathbf{c}}(\Lambda)}{\rho_{\sqrt{\Sigma_1}}(\Lambda)}\cdot\frac{\sqrt{\det(\Sigma_3)}}{\sqrt{\det(\Sigma_2)}}\notag\\
&=\frac{\rho_{\sigma,\mathbf{c}}(\Lambda)}{\rho_{r}(\mathbb{Z}^n)}\cdot\frac{\sqrt{\det(\Sigma_3)}}{\sqrt{\det(\Sigma_2)}},
\end{flalign}}where inequality $(c)$ comes from the fact that $\rho_{\sqrt{\Sigma}}(\Lambda-\mathbf{c})\leq\rho_{\sqrt{\Sigma}}(\Lambda)$.


The Lemma is proven by showing that
{\allowdisplaybreaks\begin{flalign}
\frac{\sqrt{\det(\Sigma_3)}}{\sqrt{\det(\Sigma_2)}}&\overset{(d)}{=}\sqrt{\det(\Sigma_3)}\cdot\sqrt{\det(\Sigma^{-1}_2)}\notag\\
&\overset{(e)}{=}\sqrt{\det(\Sigma_3\Sigma^{-1}_2)}\notag\\
&=\sqrt{\det(\Sigma\Sigma^{-1}_1)}\notag\\
&=\sqrt{\det\left(\frac{r^2}{\sigma^2}\cdot\mathbf{B}\mathbf{B}^T\right)}\notag\\
&=\frac{r^n}{\sigma^n}\cdot{|\det(\mathbf{B})|}.
\end{flalign}}Here, $(d)$ and $(e)$ follow from the properties of determinant that
\begin{equation}
\frac{1}{\det (\mathbf{A})}=\det(\mathbf{A}^{-1})
\end{equation}
and
\begin{equation}
\det(\mathbf{A})\det(\mathbf{B})=\det(\mathbf{AB}),
\end{equation}
respectively, for square matrices $\mathbf{A}$ and $\mathbf{B}$ of equal sizes.

\end{proof}

To satisfy the condition that $\sigma^2\mathbf{I}\succ r^2\cdot\mathbf{B}\mathbf{B}^T$, we require
\begin{equation}
\sigma>rs_1(\mathbf{B}),
\label{mp1}
\end{equation}
where $s_1(\mathbf{B})$ denotes the largest singular value of the basis $\mathbf{B}$. It is readily verified that
\begin{equation}
{s_1(\mathbf{B})\geq\max_{1\leq i \leq n}\|\mathbf{b}_i\|\geq\max_{1\leq i \leq n}\|\mathbf{\widehat{b}}_i\|.}
\label{mp2}
\end{equation}

\begin{algorithm}[t]
\caption{Independent MH Sampling Using Peikert's Algorithm}
\begin{algorithmic}[1]
\Require
$\mathbf{B}, \sigma, \mathbf{c}, \mathbf{x}_0, t_{\text{mix}}(\epsilon), \Sigma > \Sigma_1=r^2\cdot\mathbf{B}\cdot\mathbf{B}^T$;
\Ensure $\mathbf{x} \thicksim D_{\Lambda,\sigma,\mathbf{c}}$;
\State let $\mathbf{X}_0=\mathbf{x}_0$
\For {$t=$1,2,\ \ldots,\ }
\State let $\mathbf{x}$ denote the state of $\mathbf{X}_{t-1}$
\State sample $\mathbf{y}$ from the proposal distribution $q(\mathbf{y})$ in (\ref{quantity compute11a})
\State calculate the acceptance ratio $\alpha_s(\mathbf{x},\mathbf{y})$ in (\ref{quantity compute11})
\State generate a sample $u$ from the uniform density $U[0,1]$
\If {$u\leq \alpha(\mathbf{x},\mathbf{y})$}
\State let $\mathbf{X}_t=\mathbf{y}$
\Else
\State $\mathbf{X}_t=\mathbf{x}$
\EndIf
\If {$t\geq t_{\text{mix}}(\epsilon)$ }
\State output $\mathbf{x}$
\EndIf


\EndFor
\end{algorithmic}
\end{algorithm}

\begin{my1}\label{lem:PK}
For independent MH samplings based on Peikert's algorithm and on Klein's algorithm, the following relation holds:
\begin{equation}
\delta'\leq\delta.
\end{equation}
\end{my1}

\begin{proof}
According to (\ref{xxxxabc}) and (\ref{xxxxabcd}), in order to show $\delta'\leq\delta$, we need to prove that
\begin{equation}\label{eq:Peikert-rate}
\rho_{r}(\mathbb{Z}^n)\cdot\frac{\sigma^n}{r^n}\cdot\frac{1}{|\det(\mathbf{B})|}\geq\prod^n_{i=1}\rho_{\sigma_i}(\mathbb{Z}).
\end{equation}

Next, by recalling the \emph{Jacobi theta function} $\vartheta_3(\tau)=\sum^{+\infty}_{n=-\infty}e^{-\pi\tau n^2}$ with $\tau>0$, we have
\begin{equation}
\rho_{r}(\mathbb{Z})=\vartheta_3\left(\frac{1}{2\pi r^2}\right)
\end{equation}
and the left-hand side of \eqref{eq:Peikert-rate}
{\allowdisplaybreaks\begin{flalign}
&=\frac{1}{(\sqrt{2\pi}r)^n}\vartheta_3^n\left(\frac{1}{2\pi r^2}\right)\cdot(\sqrt{2\pi})^n\cdot\frac{\sigma^n}{|\det(\mathbf{B})|}\notag\\
&\overset{(f)}{=}\vartheta_3^n(2\pi r^2)\cdot(\sqrt{2\pi})^n\cdot\prod^n_{i=1}\sigma_i,
\end{flalign}}where $(f)$ utilizes the symmetry property of \emph{Theta series} for isodual lattice $\mathbb{Z}$
\begin{equation}
\vartheta_3\left(\frac{1}{\tau^2}\right)=\tau\vartheta_3(\tau^2).
\end{equation}
Moreover, as $\vartheta_3(\tau)$ is monotone decreasing with $\tau$, the following relation holds:
{\allowdisplaybreaks\begin{flalign}
\vartheta_3(2\pi r^2)&\geq\vartheta_3\left(2\pi r^2\frac{s^2_1(\mathbf{B})}{\|\widehat{\mathbf{b}}_i\|^2}\right)\notag\\
&\geq\vartheta_3\left(2\pi \frac{\sigma^2}{\|\widehat{\mathbf{b}}_i\|^2}\right)\notag\\
&=\vartheta_3(2\pi \sigma^2_i)
\end{flalign}}due to (\ref{mp1}) and (\ref{mp2}).


Hence, we finally have that the left-hand side of \eqref{eq:Peikert-rate}
{\allowdisplaybreaks\begin{flalign}
&\geq\prod^n_{i=1}(\sqrt{2\pi}\sigma_i)\cdot\vartheta_3(2\pi\sigma^2_i)\notag\\
&=\prod^n_{i=1}\vartheta_3\left(\frac{1}{2\pi\sigma^2_i}\right)\notag\\
&=\prod^n_{i=1}\rho_{\sigma_i}(\mathbb{Z}),
\end{flalign}}thus completing the proof.

\end{proof}

Similarly to independent MHK, it is easy to verify that the proposed algorithm is also uniformly ergodic.
\begin{my2}
Given $D_{\Lambda,\sigma,\mathbf{c}}$, the Markov chain induced by independent MH sampling using Peikert's algorithm converges exponentially fast:
\begin{equation}
\|P^t(\mathbf{x}, \cdot)-D_{\Lambda,\sigma,\mathbf{c}}(\cdot)\|_{TV}\leq (1-\delta')^t.
\label{eq:b65}
\end{equation}
\end{my2}

\begin{figure}[t]
\begin{center}
\includegraphics[width=3.5in]{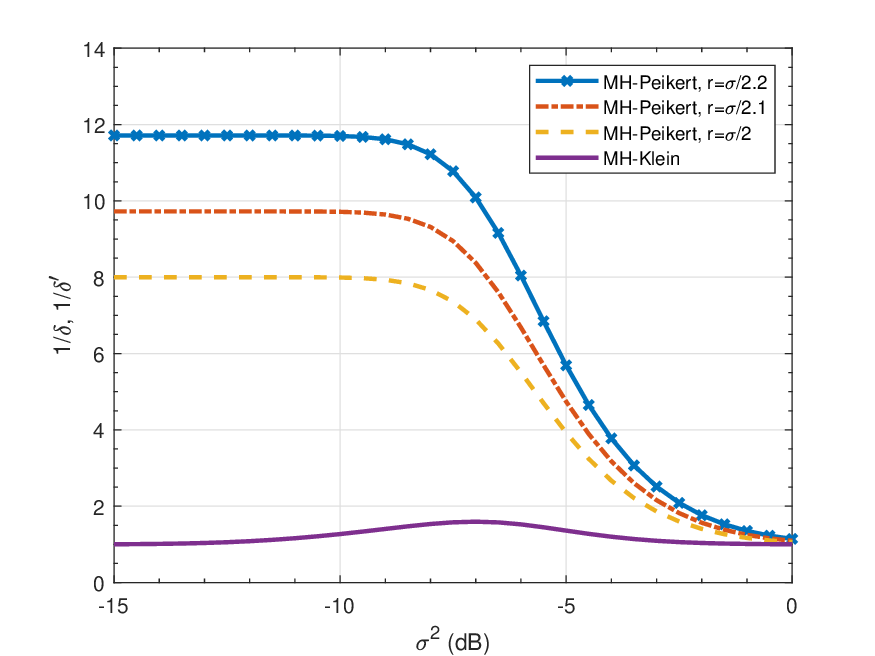}
\end{center}
\vspace{-1em}
  \caption{Comparison of $1/\delta$ and $1/\delta'$ for independent MH samplings based on Klein's and Peikert's algorithms for lattice $D_4$ with $\sigma^2=-8$dB and $\mathbf{c}=\mathbf{0}$.}
  \label{fig:PK}
\end{figure}

By Lemma \ref{lem:PK}, we can see that the independent MH sampling based on Peikert's algorithm converges slower than that based on Klein's algorithm. This is numerically confirmed in Fig. \ref{fig:PK} for checkerboard lattice $D_4$, where a comparison of the coefficients $1/\delta$ and $1/\delta'$ is given. Clearly, in the whole range of $r$, the independent MH-Peikert sampling requires more iterations than independent MHK.


\subsection{Extension to Rejection Sampling}
The classic rejection sampling is able to generate independent samples from the target distribution, but requires a normalizing constant for the application of a proposal distribution \cite{RejectionSampler}. Given the target distribution $\pi(\mathbf{x})=D_{\Lambda,\sigma,\mathbf{c}}(\mathbf{x})$, its operation consists of the following three steps:

1)\ \ \hspace{-.2em}\emph{Generate a candidate sample $\mathbf{y}$ from distribution $q(\mathbf{y})$ using Klein's algorithm or Peikert's algorithm.}

2)\ \ \hspace{-.2em}\emph{Calculate a normalizing constant $\omega_0$ such that}
\begin{equation}
\omega_0\cdot q(\mathbf{x})\geq\pi(\mathbf{x})
\label{complexdf1223245}
\end{equation}
\emph{for all $\mathbf{x}\in\mathbb{Z}^n$.}

3)\ \ \hspace{-.2em}\emph{Output $\mathbf{y}$ with probability}
\begin{equation}
\alpha=\frac{\pi(\mathbf{y})}{\omega_0\cdot q(\mathbf{y})}=\frac{\omega(\mathbf{y})}{\omega_0}
\end{equation}
\emph{and otherwise repeat.
}



Generally, rejection sampling is not directly comparable with MCMC sampling as it requires the normalizing constant $\omega_0$ for calibrating, which is not realistic in many cases of interest. Nevertheless, with a certain choice of $\omega_0$, it is possible to interpret it as a particular MCMC algorithm.

\begin{my3}
Given the target distribution $\pi(\mathbf{x})=D_{\Lambda,\sigma,\mathbf{c}}(\mathbf{x})$, the Markov chain arising from the above rejection sampler with $\omega_0\geq\pi(\mathbf{x})/q(\mathbf{x})$ for all $\mathbf{x}\in\mathbb{Z}^n$ is reversible, irreducible and aperiodic, with transition probability
{\allowdisplaybreaks\begin{flalign}
P(\mathbf{x},\mathbf{y})&=\begin{cases}q(\mathbf{y})\cdot\hspace{-.1em}\frac{w(\mathbf{y})}{w_0} \ \ \ \ \ \ \ \ \ \ \ \text{if}\ \mathbf{y}\neq\mathbf{x}, \\
1-\underset{\mathbf{z}\neq\mathbf{x}}{\sum}q(\mathbf{z})\cdot\frac{w(\mathbf{z})}{w_0}       \ \ \text{if}\ \mathbf{y}=\mathbf{x}.
       \end{cases}
\label{b3}
\end{flalign}}
\end{my3}



Clearly, the algorithm based on rejection sampling converges when the first acceptance takes place. The samples after the acceptance are naturally independently and identically distributed (i.i.d.).
Similarly to the setting in (\ref{b8}), the transition matrix $\mathbf{P}_{\text{r}}$ of this Markov chain is exactly given by
\begin{equation*}
\mathbf{P}_{\text{r}}\hspace{-0.2em}=\hspace{-0.5em}\left[\hspace{-.6em}\begin{array}{ccccc}
                                \frac{\pi_1}{\omega_0}\hspace{-.3em}+\hspace{-.3em}(\hspace{-.1em}1\hspace{-.3em}-\hspace{-.3em}\frac{1}{\omega_0}\hspace{-.2em}) & \frac{\pi_2}{\omega_0} & \frac{\pi_3}{\omega_0} & \cdots & \frac{\pi_{\infty}}{\omega_0} \\
                                   \frac{\pi_1}{\omega_0}    & \frac{\pi_2}{\omega_0}\hspace{-.3em}+\hspace{-.3em}(\hspace{-.1em}1\hspace{-.3em}-\hspace{-.3em}\frac{1}{\omega_0}\hspace{-.2em}) & \frac{\pi_3}{\omega_0} & \cdots & \frac{\pi_{\infty}}{\omega_0}\\
                                 \frac{\pi_1}{\omega_0}    &    \frac{\pi_2}{\omega_0}    &  \frac{\pi_3}{\omega_0}\hspace{-.3em}+\hspace{-.3em}(\hspace{-.1em}1\hspace{-.3em}-\hspace{-.3em}\frac{1}{\omega_0}\hspace{-.2em}) &\cdots &\frac{\pi_{\infty}}{\omega_0}\\
                                 \vdots & \vdots  & \vdots  & \ddots & \vdots\\
                                   \frac{\pi_1}{\omega_0}    &    \frac{\pi_2}{\omega_0}  &    \frac{\pi_3}{\omega_0}  & \cdots & \frac{\pi_{\infty}}{\omega_0}\hspace{-.3em}+\hspace{-.3em}(\hspace{-.1em}1\hspace{-.3em}-\hspace{-.3em}\frac{1}{\omega_0}\hspace{-.2em})
                                \end{array}\hspace{-.7em}\right]
\end{equation*}
which can be further decomposed into
\begin{equation}
\mathbf{P}_{\text{r}}=\mathbf{P}_{\text{r}}(\cdot,\cdot|\text{accept})\cdot P_{\text{accept}}+\mathbf{P}_{\text{r}}(\cdot,\cdot|\text{reject})\cdot P_{\text{reject}}
\label{b81112}
\end{equation}
where
\begin{equation}
\mathbf{P}_{\text{r}}(\cdot,\cdot|\text{accept})\hspace{-0.1em}=\hspace{-0.4em}\left[\begin{array}{ccccc}
                                \pi_1 & \pi_2 & \pi_3 & \cdots & \pi_{\infty} \\
                                \pi_1 & \pi_2 & \pi_3 & \cdots & \pi_{\infty} \\
                                \pi_1 & \pi_2 & \pi_3 & \cdots & \pi_{\infty} \\
                                 \vdots & \vdots  & \vdots  & \ddots & \vdots\\
                                \pi_1 & \pi_2 & \pi_3 & \cdots & \pi_{\infty} \\
                                \end{array}\right]
\end{equation}
\begin{equation}
\mathbf{P}_{\text{r}}(\cdot,\cdot|\text{reject})\hspace{-0.1em}=\hspace{-0.4em}\left[\begin{array}{ccccc}
                                1 & 0 & 0 & \cdots & 0 \\
                                0 & 1 & 0 & \cdots & 0 \\
                                0 & 0 & 1 & \cdots & 0 \\
                                 \vdots & \vdots  & \vdots  & \ddots & \vdots\\
                                0 & 0 & 0 & \cdots & 1 \\
                                \end{array}\right]
\end{equation}
and
\begin{equation}
\begin{cases}P_{\text{accept}}=1/\omega_0,\\
P_{\text{reject}}=1-1/\omega_0.
\end{cases}
\end{equation}
Here, $P_{\text{accept}}$ and $P_{\text{reject}}$ denote the acceptance and rejection probabilities of a new candidate in the next move.

Similarly to the analysis of independent MHK, we have the following Lemma, whose proof is omitted due to simplicity.
\begin{my1}
The eigenvalues $\eta_i$'s of the transition matrix $\mathbf{P}_{\text{r}}$ satisfy that
\end{my1}
\vspace{-1em}
\begin{equation}
1>|\eta_1|=|\eta_2|=\cdots>0
\end{equation}
\emph{with}
\begin{equation}
\eta_i=1-\frac{1}{\omega_0}
\end{equation}
\emph{for} $i=1, \ldots, \infty$.

Furthermore, we arrive at the following Theorem.
\begin{my2}
Given the invariant lattice Gaussian distribution $\pi=D_{\Lambda,\sigma,\mathbf{c}}$, the Markov chain induced by rejection sampling converges exponentially fast as
\begin{equation}
\|P^t(\mathbf{x}, \cdot)-D_{\Lambda,\sigma,\mathbf{c}}(\cdot)\|_{TV}=(1-\pi(\mathbf{x}))\cdot(\tau_1)^t,
\end{equation}
where the spectral radius $\tau_1=\eta_1=1-\frac{1}{\omega_0}$.
\end{my2}

\begin{proof}
Let $A_t$ denote the number of acceptances during consecutive $t$ moves. Then
\begin{equation*}
\|P^t(\mathbf{x}, \cdot)-D_{\Lambda,\sigma,\mathbf{c}}(\cdot)\|_{TV}=\|P^t(\mathbf{x}, \cdot|A_t=0)\cdot P(A_t=0)+
\end{equation*}
\begin{equation*}
\ \ \ \ \ \ \ \ \ \ \ \ \ \ P^t(\mathbf{x}, \cdot|A_t>0)\cdot P(A_t>0)-D_{\Lambda,\sigma,\mathbf{c}}(\cdot)\|_{TV}
\end{equation*}
\vspace{-2em}
\begin{eqnarray}
&=&\hspace{-.8em}\left\|P^t(\mathbf{x}, \cdot|A_t=0)\cdot\left(1-\frac{1}{\omega_0}\right)^t-D_{\Lambda,\sigma,\mathbf{c}}\cdot\left(1-\frac{1}{\omega_0}\right)^t\right\|_{TV}\notag\\
&=&\hspace{-.8em}\left\|[P^t(\mathbf{x}, \cdot|A_t=0)-D_{\Lambda,\sigma,\mathbf{c}}]\cdot\left(1-\frac{1}{\omega_0}\right)^t\right\|_{TV}\notag\\
&=&\hspace{-.8em}(1-D_{\Lambda,\sigma,\mathbf{c}}(\mathbf{x}))\cdot\left(1-\frac{1}{\omega_0}\right)^t\notag\\
&=&\hspace{-.8em}(1-\pi(\mathbf{x}))\cdot\tau_1^t,
\end{eqnarray}
where $P(A_t=0)=(1-1/\omega_0)^t$, $P(A_t>0)=1-(1-1/\omega_0)^t$, and $P(\mathbf{x}, \cdot|A_t>0)$ has converged to $D_{\Lambda,\sigma,\mathbf{c}}$ after the first acceptance.
\end{proof}

According to Theorem 3, the convergence rate of rejection sampling depends on the choice of the normalizing constant $\omega_0$.
Because $\omega_0\geq\pi(\mathbf{x})/q(\mathbf{x})$ for all $\mathbf{x}\in\mathbb{Z}^n$, the spectral radius $\tau_1=\eta_1$ of rejection sampling achieves the minimum when
$\omega_0=\omega_{\text{max}}(\mathbf{x})=\omega(\mathbf{x}_1)$, namely,
\begin{equation}
\tau_1=1-\frac{1}{\omega(\mathbf{x}_1)}\leq1-\delta,
\end{equation}
thus leading to
\begin{equation}
\|P^t(\mathbf{x}, \cdot)-D_{\Lambda,\sigma,\mathbf{c}}(\cdot)\|_{TV}\leq(1-\pi(\mathbf{x}))\cdot(1-\delta)^t.
\label{aaaaa1}
\end{equation}

From \eqref{b65} and (\ref{aaaaa1}), it is worth noting that only when $\omega_0=\omega(\mathbf{x}_1)$, rejection sampling and independent MH have the same convergence rate. However, the former requires the knowledge of $\omega_0$ while the latter does not.

\begin{my5}
Another algorithm for lattice Gaussian sampling based on rejection sampling was proposed in \cite{BrakerskiRejectSampling}. However, it was only concerned with values of $\sigma$ required by Klein's algorithm. Its goal is to use rejection sampling to produce exact Gaussian samples, since Klein's algorithm only approximates the target distribution. In contrast, our goal is to sample with smaller values of $\sigma$. 
The algorithm of \cite{BrakerskiRejectSampling} computes a certain normalizing constant in polynomial time and needs just a few steps on average to produce an exact sample. It is possible to extend their algorithm to smaller values of $\sigma$, but its running time will blow up.
\end{my5}

\section{Complexity of BDD}

\begin{algorithm}[t]
\caption{BDD using Independent MHK Sampling}
\begin{algorithmic}[1]
\Require
$\mathbf{B}, \sigma, \mathbf{c}, \mathbf{x}_0, t$;
\Ensure $\widehat{\mathbf{x}}$;
\State let $\widehat{\mathbf{x}}=\mathbf{x}_0$ and $\mathbf{X}_0=\mathbf{x}_0$
\For {$i=$1,\ \ldots, $t$}
\State let $\mathbf{x}$ denote the state of $\mathbf{X}_{t-1}$
\State sample $\mathbf{y}$ from the proposal distribution $q(\mathbf{x},\mathbf{y})$ in (\ref{MH proposal density})
\State calculate the acceptance ratio $\alpha(\mathbf{x},\mathbf{y})$ in (\ref{MH quantity compute})
\State generate a sample $u$ from the uniform density $U[0,1]$
\If {$u\leq \alpha(\mathbf{x},\mathbf{y})$}
\State let $\mathbf{X}_i=\mathbf{y}$ and $\mathbf{x}'=\mathbf{y}$
\If {$\|\mathbf{c}-\mathbf{B}\mathbf{x}'\|<\|\mathbf{c}-\mathbf{B}\widehat{\mathbf{x}}\|$}
\State update $\widehat{\mathbf{x}}=\mathbf{x}'$
\EndIf
\Else
\State $\mathbf{X}_i=\mathbf{x}$


\EndIf


\EndFor
\State{output $\widehat{\mathbf{x}}=\mathbf{x}'$}
\end{algorithmic}
\end{algorithm}

In this section, we apply the independent MHK sampling to BDD and analyze its complexity. The analysis for independent MH-Peikert and rejection sampling is similar, by changing the value $\delta$. As mentioned before, the decoding complexity of MCMC is evaluated by the number of Markov moves.

In MCMC, samples from the stationary distribution tend to be correlated with each other. Thus one leaves a gap, which is the mixing time $t_{\text{mix}}$, to pick up the desired independent samples (alternatively, one can run multiple Markov chains in parallel to guarantee i.i.d. samples). Therefore, we define the complexity of solving BDD by MCMC as follows.

\begin{my3}
Let $d(\Lambda, \mathbf{c})=\min_{\mathbf{x}\in \mathbb{Z}^n}\|\mathbf{Bx}-\mathbf{c}\|$ denote the Euclidean distance between the query point $\mathbf{c}$ and the lattice $\Lambda$ with basis $\mathbf{B}$, and let $\widehat{\mathbf{x}}$ be the lattice point achieving $d(\Lambda, \mathbf{c})$. The complexity (i.e., the number of Markov moves $t$) of solving BDD by MCMC is
\begin{equation}
C_{\mathrm{BDD}}\triangleq\frac{t_{\text{mix}}}{D_{\Lambda,\sigma,\mathbf{c}}(\widehat{\mathbf{x}})}.
\label{complexdf}
\end{equation}

\end{my3}

Then, $C_{\mathrm{BDD}}$ can be upper bounded by
{\allowdisplaybreaks\begin{flalign}
C_{\mathrm{BDD}}&<\log\left(\frac{1}{\epsilon}\right)\cdot\frac{1}{\delta}\cdot\frac{\rho_{\sigma, \mathbf{c}}(\mathbf{\Lambda})}{\rho_{\sigma, \mathbf{c}}(\mathbf{B}\widehat{\mathbf{x}})}\notag\\
&\leq\log\left(\frac{1}{\epsilon}\right)\cdot\frac{\prod^n_{i=1}\rho_{\sigma_{i}}(\mathbb{Z})}{\rho_{\sigma, \mathbf{c}}(\mathbf{\Lambda})}\cdot\frac{\rho_{\sigma,\mathbf{c}}(\mathbf{\Lambda})}{\rho_{\sigma,\mathbf{c}}(\mathbf{B}\widehat{\mathbf{x}})}\notag\\
&=\log\left(\frac{1}{\epsilon}\right)\cdot\frac{\prod^n_{i=1}\rho_{\sigma_{i}}(\mathbb{Z})}{\rho_{\sigma, \mathbf{c}}(\mathbf{B}\widehat{\mathbf{x}})}\notag\\
&=\log\left(\frac{1}{\epsilon}\right)\cdot C,
\label{m2}
\end{flalign}}where
\begin{equation}
C=\frac{\prod^n_{i=1}\rho_{\sigma_{i}}(\mathbb{Z})}{\rho_{\sigma, \mathbf{c}}(\mathbf{B}\widehat{\mathbf{x}})}.
\label{a9}
\end{equation}

\begin{my2}
The complexity of solving BDD by the independent MHK algorithm is bounded above as
\begin{equation}
C_{\mathrm{BDD}}\leq\log\left(\frac{1}{\epsilon}\right)\cdot 1.0039^n \cdot e^{\frac{2\pi \cdot d^2(\Lambda, \mathbf{c})}{\min_i\|\widehat{\mathbf{b}}_i\|^2}}.
\label{m2}
\end{equation}
\end{my2}

\begin{proof}
To start with, let us examine the numerator in (\ref{a9})
{\allowdisplaybreaks\begin{flalign}
\prod^n_{i=1}\rho_{\sigma_{i}}(\mathbb{Z})&=\prod_{i=1}^{n}\sum_{x_i\in\mathbb{Z}}e^{-\frac{1}{2\sigma_i^2}\|x_i\|^2}\notag\\
&{=}\prod_{i=1}^{n}\vartheta_3(\|\widehat{\mathbf{b}}_i\|^2/2\pi\sigma^2)\hspace{-0.5em}
\label{a1}
\end{flalign}}
where we apply the {Jacobi theta function} $\vartheta_3$ \cite{ConwayandSloane}.

By substituting (\ref{a1}) to (\ref{a9}), the complexity $C$ is upper bounded as
\begin{equation}
C \leq {\prod_{i=1}^{n}\vartheta_3(\|\widehat{\mathbf{b}}_i\|^2/2\pi\sigma^2)}\cdot{ e^{\frac{1}{2\sigma^2}\|\mathbf{B}\widehat{\mathbf{x}}-\mathbf{c}\|^2}}.
\label{b22}
\end{equation}

Now, let us recall some facts about Jacobi theta function $\vartheta_3(\tau)$. $\vartheta_3(\tau)$ is monotonically decreasing with $\tau$, and particularly
\begin{equation}
\lim_{\tau\rightarrow\infty}\inf\vartheta_3(\tau)=1.
\label{b17}
\end{equation}
By simple calculation, we can get that
\begin{equation}
\vartheta_3(2)=\sum^{+\infty}_{n=-\infty}e^{-2\pi n^2}=\frac{\sqrt[4]{6\pi+4\sqrt{2}\pi}}{2\Gamma(\frac{3}{4})}=1.0039,
\label{b18}
\end{equation}
where $\Gamma(\cdot)$ stands for the \emph{Gamma function}. Clearly, if
\begin{equation}
\frac{\min_{1\leq i\leq n}\|\widehat{\mathbf{b}}_i\|^2}{2\pi\sigma^2}\geq 2
\label{b20}
\end{equation}
it turns out that the following term
\begin{equation}
\prod_{i=1}^{n}\vartheta_3(\|\widehat{\mathbf{b}}_i\|^2/2\pi\sigma^2)\leq\vartheta_3^n(2)=1.0039^n
\end{equation}
is rather small even for values of $n$ up to hundreds (e.g., $1.0039^{100}=1.4467$). The key point here is that the pre-exponential factor is close to 1. For better accuracy, $\vartheta_3(3)=1.00037$ (or $\vartheta_3(4)$ etc.) can be applied so that $1.00037^{1000}=1.4476$. More options about $\vartheta_3$ can be found in Table I.

Therefore, if $\sigma$ satisfies the condition (\ref{b20}), namely
\begin{equation}
\sigma\leq\min_{1\leq i\leq n}\|\widehat{\mathbf{b}}_i\|/(2\sqrt{\pi}),
\end{equation}
then we have
\begin{equation}
C\leq 1.0039^n \cdot e^{\frac{1}{2\sigma^2}\|\mathbf{B}\widehat{\mathbf{x}}-\mathbf{c}\|^2}.
\label{d2}
\end{equation}
Setting $\sigma=\min_{i}\|\widehat{\mathbf{b}}_i\|/(2\sqrt{\pi})$, we finally arrive at the following result
{\allowdisplaybreaks\begin{flalign}
C_{\text{BDD}}&\leq\log\left(\frac{1}{\epsilon}\right)\cdot 1.0039^n \cdot e^{\frac{2\pi}{\min_i\|\widehat{\mathbf{b}}_i\|^2}\|\mathbf{B}\widehat{\mathbf{x}}-\mathbf{c}\|^2},
\label{d21}
\end{flalign}}completing the proof.
\end{proof}

Let us highlight the significance of lattice reduction. Lattice reduction is able to significantly improve $\min_i\|\widehat{\mathbf{b}}_i\|$ while reducing $\max_i\|\widehat{\mathbf{b}}_i\|$ \cite{LLLoriginal}. Therefore, increasing $\min_i\|\widehat{\mathbf{b}}_i\|$ will significantly decrease the complexity shown above.


%
%

\renewcommand{\arraystretch}{1.8}
\begin{table}[t]
\begin{center}
\caption{Values of $\vartheta_3$.}
\label{tab:TCQvsPL}
\begin{tabular}{|c||c||c||c|}\hline
$\vartheta_3(1)$ & $\sum^{+\infty}_{n=-\infty}e^{-1\pi n^2}$ &  $\frac{\sqrt[4]{\pi}}{\Gamma(\frac{3}{4})}$ & 1.087 \\\hline
$\vartheta_3(2)$ & $\sum^{+\infty}_{n=-\infty}e^{-2\pi n^2}$  & $\frac{\sqrt[4]{6\pi+4\sqrt{2}\pi}}{2\Gamma(\frac{3}{4})}$ & 1.0039   \\\hline
$\vartheta_3(3)$ & $\sum^{+\infty}_{n=-\infty}e^{-3\pi n^2}$ &  $\frac{\sqrt[4]{27\pi+18\sqrt{3}\pi}}{3\Gamma(\frac{3}{4})}$ & 1.00037 \\\hline
$\vartheta_3(4)$ & $\sum^{+\infty}_{n=-\infty}e^{-4\pi n^2}$ &  $\frac{\sqrt[4]{8\pi}+2\sqrt[4]{\pi}}{4\Gamma(\frac{3}{4})}$ & 1.0002 \\\hline
$\vartheta_3(5)$ & $\sum^{+\infty}_{n=-\infty}e^{-5\pi n^2}$ &  $\frac{\sqrt[4]{225\pi+100\sqrt{5}\pi}}{5\Gamma(\frac{3}{4})}$ & 1.0001 \\\hline
\end{tabular}
\end{center}
\end{table}

\begin{my5}
In fact, such an analysis also holds for Klein's algorithm, where the probability of sampling $\mathbf{x}$ follows a Gaussian-like distribution\cite{Klein}
\begin{equation}
P(\mathbf{x})\geq \frac{e^{-\frac{1}{2\sigma^2}\|\mathbf{Bx}-\mathbf{c}\|^2}}{\prod_{i=1}^{n}\vartheta_3(\|\widehat{\mathbf{b}}_i\|^2/2\pi \sigma^2)}.
\label{b11}
\end{equation}
Klein chose $\sigma=\min_i\|\mathbf{\widehat{b}}_i\|/\sqrt{2\log n}$, which corresponds to $O(n^{d^2(\Lambda, \mathbf{c})/\min_i\|\mathbf{\widehat{b}}_i\|^2})$ complexity. Here, we have shown that the decoding complexity can be further reduced to $O(e^{d^2(\Lambda, \mathbf{c})/\min_i^2\|\widehat{\mathbf{b}}_i\|})$, by setting $\sigma=\min_{i}\|\widehat{\mathbf{b}}_i\|/(2\sqrt{\pi})$. With the help of HKZ reduction, $\min_{i}\|\widehat{\mathbf{b}}_i\| \geq \frac{1}{n}\lambda_1(\Lambda)$ \cite{lagarias}. Thus, Klein's algorithm allows to solve the $\eta$-BDD with $\eta=O(1/n)$ in polynomial time, while our result shown in (\ref{d21}) improves it to $\eta=O(\sqrt{\log n}/n)$.
\end{my5}


According to \eqref{m2}, we have
\begin{equation}
d(\Lambda, \mathbf{c})=\sqrt{\frac{1}{2\pi}\cdot\ln\frac{C_{\mathrm{BDD}}}{a}}\cdot\min_{1\leq i\leq n}\|\widehat{\mathbf{b}}_i\|.
\label{d5}
\end{equation}
where $a=\log\left(\frac{1}{\epsilon}\right)\cdot 1.0039^n \approx \log\left(\frac{1}{\epsilon}\right)$.
Clearly, the decoding radius increases with $C_{\mathrm{BDD}}$, implying a flexible trade-off between the decoding performance and complexity. In addition, the significance of lattice reduction can be seen due to an increased value of $\min_{i}\|\widehat{\mathbf{b}}_i\|$.

\section{Trapdoor Sampling}

The core technique underlying GPV's signature scheme is discrete Gaussian sampling over a trapdoor lattice \cite{Trapdoor}. Its security crucially relies on the property that the output distribution of discrete Gaussian sampling is oblivious to any particular basis used in the sampling process, therefore preventing leakage of the private key. The original GPV signature scheme was based on Klein's algorithm, which was subsequently extended to Peikert's algorithm \cite{Peikert10} (see also \cite[Chap. 6]{Prest} for sampling over structured lattices). In fact, any good Gaussian sampling algorithms can be applied to GPV signatures. In this Section, we demonstrate the security advantage of MCMC in GPV signatures, thanks to smaller parameters it can reach.

Firstly, we provide a high-level introduction to the GPV signature (see \cite{Trapdoor,Peikert10} for details). In key generation, one generates a hard public basis for a random lattice $\Lambda$, together with a short private basis of $\Lambda$. The public basis serves as the public key, while the private basis serves as the private key. Given a message $\mathbf{m}$ (or rather a digest of $\mathbf{m}$), one uses the private basis to sample a point $\mathbf{x}$ from $D_{\Lambda+\mathbf{m},\sigma}$ with parameter $\sigma$. The signature of $\mathbf{m}$ is $\mathbf{x}$. The verifier checks that $\mathbf{x}$ is short and that $\mathbf{x}-\mathbf{m} \in \Lambda$ using the public basis.

It is shown in \cite{Trapdoor} that the security of GPV signing can be reduced to the hardness of the inhomogeneous short integer solution (ISIS) problem\footnote{In the language of coding theory, this is to find a short vector in a coset of a linear code.} with approximation factor $\sqrt{n}\sigma$. Therefore, the width $\sigma$ is the most important property of a discrete Gaussian sampler in this context.

Obviously, there is a tradeoff between security and running time in trapdoor sampling with MCMC. A small parameter $\sigma$ gives higher security, but require longer running time. Next, we examine the impact of decreasing $\sigma$ on the mixing time. Again, we focus on the independent MHK algorithm. Recall its the mixing time is proportional to
\begin{equation}
\frac{1}{\delta}=\frac{\prod^n_{i=1}\rho_{\sigma_i}(\mathbb{Z})}{\rho_{\sigma, \mathbf{c}}(\mathbf{\Lambda})}.
\label{eq:mixing}
\end{equation}

\begin{figure}[t]
\vspace{-2em}
\begin{center}
\hspace{-1em}\includegraphics[width=3.5in]{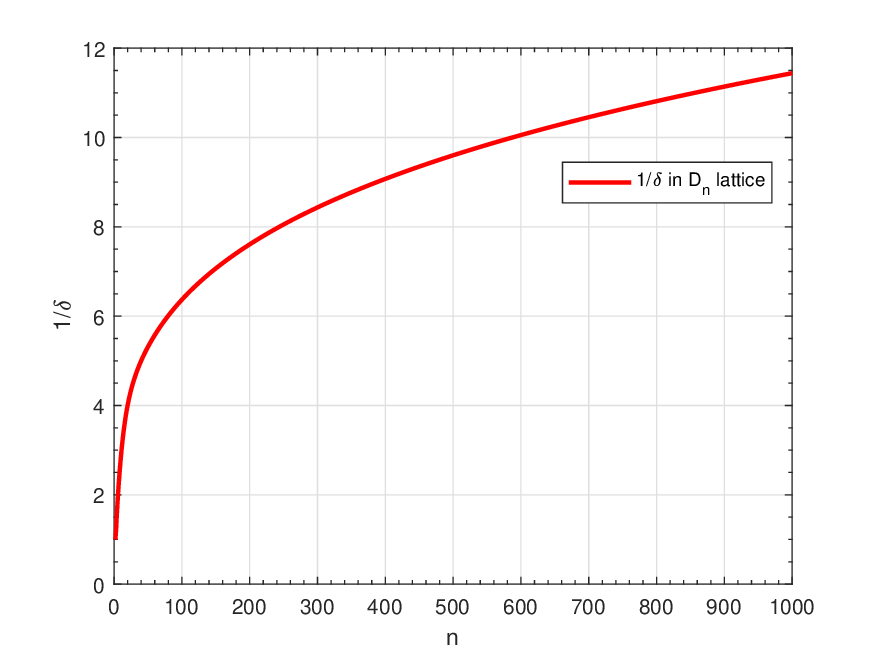}
\end{center}
\vspace{-1em}
  \caption{$\frac{1}{\delta}$ as a function of $n$ for lattice $D_n$ with $\mathbf{c}=0$ and $\sigma^2 = -8$ dB.}
  \label{fig:D4}
\end{figure}

Our intuition here is that if a good basis is available (as in the case of trapdoor sampling), then $\frac{1}{\delta}$ will not blow up as $n$ grows. To give an impression, Fig. \ref{fig:D4} shows $\frac{1}{\delta}$ as a function of $n$ for checkerboard lattice $D_n$ with $\mathbf{c}=\mathbf{0}$ and $\sigma^2 = -8$ dB, using its well-known basis \cite[p.117, (86)]{ConwayandSloane}. It is seen that $\frac{1}{\delta}$ merely grows to 12 for $n$ up to 1000.

What if $\mathbf{c}\neq \mathbf{0}$? Then the denominator of \eqref{eq:mixing} can be unpredictable in general. Fortunately, it can be bounded if $\sigma$ is above the smoothing parameter. Recall that for a lattice $\Lambda$ and for $\varepsilon > 0$, the smoothing parameter\footnote{Note again the difference from the definition in~\cite{MicciancioGaussian}, where~$\sigma$ is scaled by a constant factor $\sqrt{2\pi}$.}
$\eta_{\varepsilon}(\Lambda)$ is defined as the smallest $\sigma>0$ such that
$\sum_{{\bf{x}^*}\in \Lambda^* \setminus \{\mathbf{0}\}} e^{-2\pi^2
\sigma^2\|{\bf{x}^*}\|^2}\leq \varepsilon$.
If $\varepsilon < 1$, we have $\frac{\rho_{\sigma, \mathbf{c}}(\mathbf{\Lambda})}{(\sqrt{2\pi}\sigma)^n} \in \frac{1}{\mathrm{Vol}(\Lambda)}[1-\varepsilon, 1+\varepsilon]$, $\forall \mathbf{c}$.

Here, we are concerned with the parameter region $\sigma \in [\eta_{\varepsilon}(\Lambda), \sqrt{\omega({\log}\ n)}\cdot{\max}\|\mathbf{\widehat{b}}_i\|]$, below GPV's parameter \cite{Trapdoor} but above the smoothing parameter. This is because we anticipate only moderate growth in mixing time but significant increase of security for values of $\sigma$ just below GPV's parameter.

Let $I$ denote the subset of indexes $i$ with $\sqrt{2\pi}\sigma_i>1$ (i.e., $\sqrt{2\pi}\sigma>\|\widehat{\mathbf{b}}_i\|$), $i\in \{1, 2, \ldots, n\}$, $|I|=m$. It is not difficult to derive the following bound, similarly to \cite[Proposition 4]{ZhengWangTIT15}:
\begin{eqnarray}
\frac{1}{\delta}&=&\frac{\prod^n_{i=1}\vartheta_{3}(\frac{1}{{2\pi}\sigma_i^2})}{\rho_{\sigma, \mathbf{c}}(\mathbf{\Lambda})}\notag\\
&\in&\frac{\prod^n_{i=1}\sqrt{2\pi}\sigma_i\vartheta_3({2\pi}\sigma_i^2)}{(\sqrt{2\pi}\sigma)^n/\mathrm{Vol(\Lambda)}}[1-2\varepsilon, 1+2\varepsilon]\notag\\
&=&\prod^n_{i=1}\vartheta_3({2\pi}\sigma_i^2)[1-2\varepsilon, 1+2\varepsilon]\label{eq:delta-bound}\\
&\leq&\vartheta_3(1)^{m}\cdot \prod_{i\notin I}\frac{2}{\sqrt{2\pi}\sigma_i}\cdot (1+2\varepsilon)\notag
\end{eqnarray}
where we use the identity $\vartheta_3\left(\frac{1}{\tau^2}\right)=\tau\vartheta_3(\tau^2)$ and assume $\varepsilon < 1/2$ in the second step, and $\vartheta_3(\tau)\leq 1+\sqrt{\frac{1}{\tau}}$ in the last step.

Particularly, if $\sqrt{2\pi}\sigma\geq \sqrt{\alpha}\max_{1\leq i\leq n}\|\widehat{\mathbf{b}}_i\|$ for some $\alpha\geq 1$, we derive
\begin{equation}
\frac{1}{\delta}\leq\vartheta_3(\alpha)^{n}(1+2\varepsilon).
\end{equation}

Again, our key observation is that the mixing time $\vartheta_3(\alpha)^{n}$ grows rather slowly for values of $\alpha$ that are not too small. For example, when $\alpha=2$, we have $\vartheta_3(2)^{n} = 1.0039^{1000} = 49$ for $n=1000$. This means that with roughly $49$ iterations, our MCMC sampler is able to reduce the parameter from $\sqrt{\omega(\log n)}\max_{1\leq i\leq n}\|\widehat{\mathbf{b}}_i\|$ to $\frac{1}{\sqrt{\pi}}\max_{1\leq i\leq n}\|\widehat{\mathbf{b}}_i\|$. Therefore, if one is willing to use a slower signature scheme in return for higher security, MCMC offers such an option.

\begin{figure}[t]
\vspace{-2em}
\begin{center}
\hspace{-1em}\includegraphics[width=3.5in]{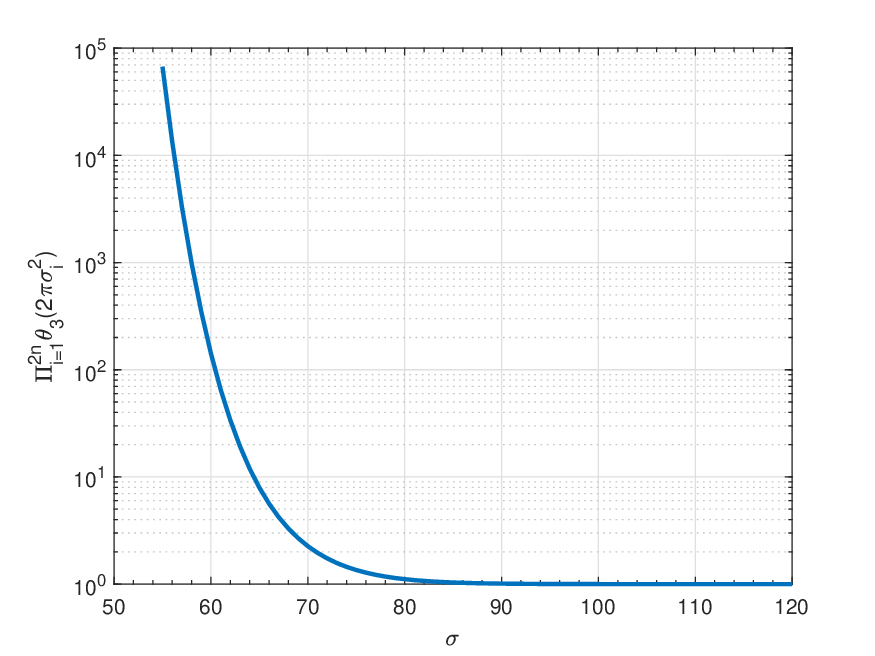}
\end{center}
\vspace{-1em}
  \caption{$\prod_{i=1}^{2n}\theta_3(2\pi\sigma_i^2)$ as a function of $\sigma$ for an NTRU lattice with $n=512$.}
  \label{fig:NTRU}
\end{figure}

\newtheorem{exap}{Example}
\begin{exap}[FALCON]
FALCON \cite{FALCON} is a GPV signature scheme instantiated by NTRU lattices. Let $m$ be a power of two, $n=\varphi(m)$ where $\varphi(\cdot)$ is Euler's totient function, $q\in \mathbb{N}$. The secret key consists of two polynomials $f$ and $g$ in ring $R=\mathbb{Z}[x]/(x^n+1)$ where $f$ is invertible. Find $G$ and $F$ such that
\[
fG-gF=q \mod x^n+1.
\]
The NTRU lattice of dimension $2n$ is generated by the private basis
\[
\mathbf{B} = \left(
               \begin{array}{cc}
                 \mathcal{C}(g) & -\mathcal{C}(f) \\
                 \mathcal{C}(G) & -\mathcal{C}(F) \\
               \end{array}
             \right)^T
\]
where $\mathcal{C}(\cdot)$ denotes an $n\times n$ nega-cyclic matrix whose first row consists of the coefficients of a polynomial. The public basis is given by
\[
\mathbf{A} = \left(
               \begin{array}{cc}
                 -\mathcal{C}(h) & \mathbf{I}_n \\
                 q\mathbf{I}_n & \mathbf{O}_n \\
               \end{array}
             \right)^T
\]
where $h=g/f\mod q$. Both bases $\mathbf{B}$ and $\mathbf{A}$ generate the same lattice
\[
\Lambda = \{(\mathbf{u},\mathbf{v})\in R^2|\mathbf{u}+\mathbf{vh} = 0 \mod q\}
\]

We consider the parameters $n=512$ and $q=12289$ in FALCON. The coefficients of polynomials $f$ and $g$ are randomly sampled from $D_{\mathbb{Z},4.05}$. For a particular instance randomly generated, we find $\max_{1\leq i\leq n}\|\widehat{\mathbf{b}}_i\|=127$. In Fig. \ref{fig:NTRU}, we show as a function of $\sigma$ the term $\prod_{i=1}^{2n}\theta_3(2\pi\sigma_i^2)$ in \eqref{eq:delta-bound}, which characterizes the complexity $1/\delta$ above the smoothing parameter. It is seen that MCMC is able to significantly reduce the parameter $\sigma$, with quite moderate increase in complexity. Specifically, the term $\prod_{i=1}^{2n}\theta_3(2\pi\sigma_i^2)$ merely grows to about 20, even if $\sigma$ is halved relative to $\max_{1\leq i\leq n}\|\widehat{\mathbf{b}}_i\|$. Recall that GPV sampling requires $\sigma = \sqrt{\omega(\log\ n)}\cdot\max_{1\leq i \leq n}\|\mathbf{\widehat{b}}_i\|$.

\end{exap}

Note that it is possible for MCMC to incorporate the fast Fourier sampler \cite{FALCON}, which would speed up the sampling process for structured lattices. The security levels of various samplers have been evaluated in \cite[Chap. 6]{Prest}. We leave evaluation of the concrete security of MCMC samplers to future work.

\section{Multiple-Try Metropolis-Klein Algorithm}
In this section, the independent multiple-try Metropolis-Klein (MTMK) algorithm is proposed to enhance the mixing. We firstly prove its validity and then show its uniform ergodicity with an improved convergence rate.

\subsection{Multiple-Try Metropolis Method}
Rather than directly generating the state candidate $\mathbf{y}$ from the proposal distribution $q(\mathbf{x}, \mathbf{y})$, the multiple-try Metropolis (MTM) method selects $\mathbf{y}$ among a set of i.i.d. trial samples from $q(\mathbf{x}, \mathbf{y})$, which significantly expands the searching region of proposals \cite{LiuMultipleTry}. In particular, the MTM method consists of the following steps:

1)\ \ \hspace{-.2em}\emph{Given the current state $\mathbf{X}_t=\mathbf{x}$, draw $k$ i.i.d. state candidates $\mathbf{y}_1, \ldots, \mathbf{y}_k$ from the proposal distribution $q(\mathbf{x},\mathbf{y})$}.

2)\ \ \hspace{-.2em}\emph{Select $\mathbf{y}=\mathbf{y}_c$ among \{$\mathbf{y}_1, \ldots, \mathbf{y}_k$\}} \emph{with probability proportional to the weight}
\begin{equation}
\omega(\mathbf{y}_i, \mathbf{x})=\pi(\mathbf{y}_i)q(\mathbf{y}_i, \mathbf{x})\lambda(\mathbf{y}_i, \mathbf{x}), \ \ i=1, \ldots, k,
\label{certain weight}
\end{equation}
\emph{where} $\lambda(\mathbf{y}, \mathbf{x})$ \emph{is a nonnegative symmetric function of} $\mathbf{y}$ \emph{and} $\mathbf{x}$ \emph{defined initially}.

3)\ \ \hspace{-.2em}\emph{Draw $k-1$ i.i.d. reference candidates $\mathbf{x}_1, \ldots, \mathbf{x}_{k-1}$ from the proposal distribution $q(\mathbf{y},\mathbf{x})$ and let $\mathbf{x}_k=\mathbf{x}$}.

4)\ \ \hspace{-.2em}\emph{Accept $\mathbf{y}=\mathbf{y}_c$ as the state of $\mathbf{X}_{t+1}$, i.e., $\mathbf{X}_{t+1}=\mathbf{y}$ with probability}
\begin{equation}
\alpha_{\text{MTM}}=\text{min}\left\{1, \frac{\omega(\mathbf{y}_1, \mathbf{x})+\ldots+\omega(\mathbf{y}_k, \mathbf{x})}{\omega(\mathbf{x}_1, \mathbf{y})+\ldots+\omega(\mathbf{x}_k, \mathbf{y})}\right\},
\end{equation}
\emph{otherwise, with probability} $1-\alpha_{\text{MTM}}$, \emph{let} $\mathbf{X}_{t+1}=\mathbf{X}_{t}=\mathbf{x}$.

By exploring the search region more thoroughly, an improvement of convergence can be achieved by MTM. Based on a number of trial samples generated from the proposal distribution, the Markov chain enjoys a large step-size jump within every single move without lowering the acceptance rate. It should be noticed that the $k-1$ reference samples $\mathbf{x}_{i}$'s are involved only for the validity of MTM by satisfying the detailed balance condition \cite{LiuMultipleTry}
\begin{equation}
\pi(\mathbf{x})P(\mathbf{x},\mathbf{y})=\pi(\mathbf{y})P(\mathbf{y},\mathbf{x}).
\label{c21}
\end{equation}

\newcounter{TempEqCnt}                         
\setcounter{TempEqCnt}{\value{equation}} 
\setcounter{equation}{85}
\begin{figure*}
\begin{eqnarray}
p(\mathbf{y}_c|\mathbf{x},c)\hspace{-.8em}&=&\hspace{-.8em}\sum_{\mathbf{y}_{1:c-1}\in\mathbb{Z}^n}\sum_{\mathbf{y}_{c+1:k}\in\mathbb{Z}^n}\left\{\left[\prod_{j=1}^kq(\mathbf{x},\mathbf{y}_j)\right]\cdot\frac{\omega(\mathbf{y}_c)}{\sum_{i=1}^k\omega(\mathbf{y}_i)}\cdot\text{min}\left\{1, \frac{\omega(\mathbf{y}_c)+\sum_{j=1,j\neq c}^k\omega(\mathbf{y}_j)}{\omega(\mathbf{x})+\sum_{j=1,j\neq c}^{k}\omega(\mathbf{y}_j)}\right\}\right\}\notag\\
\hspace{-.8em}&=&\hspace{-.8em}q(\mathbf{y}_c)\cdot\omega(\mathbf{y}_c)\cdot\hspace{-1em}\sum_{\mathbf{y}_{1:c-1}\in\mathbb{Z}^n}\sum_{\mathbf{y}_{c+1:k}\in\mathbb{Z}^n}\left\{\left[\prod_{j=1,j\neq c}^kq(\mathbf{y}_j)\right]\cdot\text{min}\left\{\frac{1}{\omega(\mathbf{y}_c)+\sum_{j=1,j\neq c}^{k}\omega(\mathbf{y}_j)}, \frac{1}{\omega(\mathbf{x})+\sum_{j=1,j\neq c}^{k}\omega(\mathbf{y}_j)}\right\}\right\}\notag\\
\hspace{-.8em}&=&\hspace{-.8em}\pi(\mathbf{y}_c)\cdot\text{min}\hspace{-0.2em}\left\{\hspace{-0.2em}\sum_{\mathbf{y}_{1:c-1}\in\mathbb{Z}^n}\hspace{-0.2em}\sum_{\mathbf{y}_{c+1:k}\in\mathbb{Z}^n}\hspace{-0.5em}\left\{\hspace{-0.2em}\frac{\prod_{j=1,j\neq c}^kq(\mathbf{y}_j)}{\omega(\mathbf{y}_c)+\sum_{j=1,j\neq c}^{k}\omega(\mathbf{y}_j)}\hspace{-0.2em}\right\}\hspace{-0.2em}, \sum_{\mathbf{y}_{1:c-1}\in\mathbb{Z}^n}\hspace{-0.2em}\sum_{\mathbf{y}_{c+1:k}\in\mathbb{Z}^n}\hspace{-0.5em}\left\{\hspace{-0.2em}\frac{\prod_{j=1,j\neq c}^kq(\mathbf{y}_j)}{\omega(\mathbf{x})+\sum_{j=1,j\neq c}^{k}\omega(\mathbf{y}_j)}\hspace{-0.2em}\right\}\hspace{-0.2em}\right\}
\label{c2}
\end{eqnarray}
\hrulefill
\end{figure*}
\setcounter{equation}{\value{TempEqCnt}}

Clearly, the efficiency of MTM relies on the number of trial samples $k$ while the traditional MH sampling is a special case with $k=1$. Similar to MH sampling, there is considerable flexibility in the choice of the proposal distribution $q(\mathbf{x}, \mathbf{y})$ in MTM \cite{LucaFlexibility}. Actually, it is even possible to use different proposal distributions to generate trial samples without altering the ergodicity of the Markov chain \cite{CasarinInteracting}. Meanwhile, the nonnegative symmetric function $\lambda(\mathbf{x}, \mathbf{y})$ in (\ref{certain weight}) is also flexible, where the only requirement is that $\lambda(\mathbf{x}, \mathbf{y})>0$ whenever $q(\mathbf{x}, \mathbf{y})>0$.

\begin{algorithm}[t]
\caption{Independent Multiple-try Metropolis-Klein Sampling Decoder}
\begin{algorithmic}[1]
\Require
$\mathbf{B}, \sigma, \mathbf{c}, \mathbf{x}_0, t$;
\Ensure $\mathbf{x} \thicksim D_{\Lambda,\sigma,\mathbf{c}}$;
\State let $\widehat{\mathbf{x}}=\mathbf{x}^0$ and $\mathbf{X}_0=\mathbf{x}_0$
\For {$i=$1,\ \ldots, $t$}
\State let $\mathbf{x}$ denote the state of $\mathbf{X}_{t-1}$
\State sample $k$ trial samples $\mathbf{y}_1\ldots\mathbf{y}_k$ from $q(\mathbf{x},\mathbf{y})$ in (\ref{m1})
\State select $\mathbf{y}=\mathbf{y}_c$ from $\mathbf{y}_1\ldots\mathbf{y}_k$ based on $\omega(\mathbf{y}_i)$ in (\ref{certain weightts})
\State calculate the acceptance ratio $\alpha(\mathbf{x},\mathbf{y})$ in (\ref{c1})
\State generate a sample $u$ from the uniform density $U[0,1]$
\If {$u\leq \alpha(\mathbf{x},\mathbf{y})$}
\State let $\mathbf{X}_i=\mathbf{y}$ and $\mathbf{x}'=\mathbf{y}$
\If {$\|\mathbf{c}-\mathbf{B}\mathbf{x}'\|<\|\mathbf{c}-\mathbf{B}\widehat{\mathbf{x}}\|$}
\State update $\widehat{\mathbf{x}}=\mathbf{x}'$
\EndIf
\Else
\State $\mathbf{X}_i=\mathbf{x}$
\EndIf


\EndFor
\State{output $\widehat{\mathbf{x}}=\mathbf{x}'$}
\end{algorithmic}
\end{algorithm}

\subsection{The Proposed Algorithm}
With the great flexibility offered by $q(\mathbf{x}, \mathbf{y})$ and $\lambda(\mathbf{x}, \mathbf{y})$, we now propose the independent multiple-try Metropolis-Klein (MTMK) algorithm, which is described by the following steps:

1)\ \ \hspace{-.2em}\emph{Given the current state $\mathbf{X}_t=\mathbf{x}$, use Klein's algorithm to draw $k$ i.i.d. state candidates $\mathbf{y}_1, \ldots, \mathbf{y}_k$ from the independent proposal distribution in (\ref{MH proposal density})}
\begin{equation}
q(\mathbf{x},\mathbf{y})=\prod^n_{i=1} P(y_{n+1-i}|\overline{\mathbf{y}}_{[-(n+1-i)]})=q(\mathbf{y}).
\label{m1}
\end{equation}

2)\ \ \hspace{-.2em}\emph{Let $\lambda(\mathbf{x},\mathbf{y})=[q(\mathbf{x},\mathbf{y})q(\mathbf{y},\mathbf{x})]^{-1}=[q(\mathbf{y})q(\mathbf{x})]^{-1}$. Then select $\mathbf{y}=\mathbf{y}_c$ among \{$\mathbf{y}_1, \ldots, \mathbf{y}_k$\}} \emph{with probability proportional to the importance weight}
\begin{equation}
\omega(\mathbf{y}_i, \mathbf{x})=\frac{\pi(\mathbf{y}_i)}{q(\mathbf{y}_i)}=\omega(\mathbf{y}_i), \ \ i=1, \ldots, k.
\label{certain weightts}
\end{equation}

3)\ \ \hspace{-.2em}\emph{Accept $\mathbf{y}=\mathbf{y}_c$ as the state of $\mathbf{X}_{t+1}$ with acceptance rate}
\begin{equation}
\alpha_{\text{MTM}}=\text{min}\left\{1, \frac{\omega(\mathbf{y}_c)+\sum_{j=1,j\neq c}^k\omega(\mathbf{y}_j)}{\omega(\mathbf{x})+\sum_{j=1,j\neq c}^{k}\omega(\mathbf{y}_j)}\right\},
\label{c1}
\end{equation}
\emph{otherwise, with probability} $1-\alpha_{\text{MTM}}$, \emph{let} $\mathbf{X}_{t+1}=\mathbf{X}_{t}=\mathbf{x}$.

In the proposed algorithm, the basic formulation of MTM is modified in three aspects. First, Klein's algorithm is applied to generate trial state candidates from the independent proposal distribution $q(\mathbf{x},\mathbf{y})=q(\mathbf{y})$. Then, by setting $\lambda(\mathbf{x},\mathbf{y})=[q(\mathbf{x},\mathbf{y})q(\mathbf{y},\mathbf{x})]^{-1}$, $\omega(\mathbf{x},\mathbf{y})$ becomes the \emph{importance weight} of $\mathbf{x}$ that we have defined in (\ref{importance weight}). Finally and interestingly, thanks to the independent proposals, the generation of reference samples $\mathbf{x}_{i}$'s can be removed without changing the ergodicity of the chain.

In the case of independent proposals, because both the trial samples $\mathbf{y}_{i}$'s and the reference samples $\mathbf{x}_{i}$'s are generated independently from the identical distribution $q(\cdot)$, the generation of reference samples can be greatly simplified by just setting $\mathbf{x}_i=\mathbf{y}_i$ for $i=1,\ldots, c-1, c+1, \ldots, k$ and $\mathbf{x}_c=\mathbf{x}$. Actually, with the same arguments, the trial samples generated in the previous Markov moves can also be used by $\mathbf{x}_i$ \cite{MartinoIMTM}.

It is well known that a Markov chain which is irreducible and aperiodic will be ergodic if the detailed balance condition is satisfied \cite{mixingtimemarkovchain}. Since irreducible and aperiodic are easy to verify, we show the validity of the proposed algorithm by demonstrating the detailed balance condition.
\begin{my2}
Given the target lattice Gaussian distribution $D_{\Lambda, \sigma, \mathbf{c}}$, the Markov chain induced by the independent MTMK algorithm is ergodic.
\end{my2}
\begin{proof}
To start with, let us specify the transition probability $P(\mathbf{x}, \mathbf{y})$ of the underlying Markov chain. For ease of presentation, we only consider the case of $\mathbf{x}\neq\mathbf{y}$, since the case $\mathbf{x}=\mathbf{y}$ is trivial. The transition probability $P(\mathbf{x}, \mathbf{y})$ can be expressed as
\begin{equation}
P(\mathbf{x}, \mathbf{y}=\mathbf{y}_c)=\sum_{i=1}^k p(\mathbf{y}_c|\mathbf{x},c=i).
\end{equation}
Here, $p(\mathbf{y}_c|\mathbf{x},c=i)$ represents the probability of accepting $\mathbf{y}=\mathbf{y}_c$ as the new state of $\mathbf{X}_{t+1}$ given the previous one $\mathbf{X}_{t}=\mathbf{x}$ when the $c$th candidate among $\mathbf{y}_i$'s is selected. Moreover, as $\mathbf{y}_i$ is exchangeable and independent, it follows that $p(\mathbf{y}_i|\mathbf{x},i)=p(\mathbf{y}_j|\mathbf{x},j)$ by symmetry, namely,
\begin{equation}
P(\mathbf{x}, \mathbf{y}=\mathbf{y}_c)=k\cdot p(\mathbf{y}_c|\mathbf{x},c).
\label{c33}
\end{equation}

In contrast to MH algorithms, the generation of the state candidate $\mathbf{y}=\mathbf{y}_c$ for Markov move $\mathbf{X}_{t+1}$ in MTM actually follows a distribution formed by $q(\mathbf{x},\mathbf{y})$ and $\omega(\mathbf{y}, \mathbf{x})$ together \cite{LiuMultipleTry}. More precisely, $p(\mathbf{y}_c|\mathbf{x},c)$ can be further expressed as (\ref{c2}), where the terms inside the sum correspond to $q(\mathbf{x},\mathbf{y})$, $\omega(\mathbf{y}, \mathbf{x})$ and $\alpha$ respectively.

\setcounter{equation}{86}
From (\ref{c2}), it is straightforward to verify the term $\pi(\mathbf{x}) p(\mathbf{y}_c|\mathbf{x},c)$ is symmetric in $\mathbf{x}$ and $\mathbf{y}_c$, namely
\begin{equation}
\pi(\mathbf{x}) p(\mathbf{y}_c|\mathbf{x},c)=\pi(\mathbf{y}_c) p(\mathbf{x}|\mathbf{y}_c,c).
\end{equation}
Then, by simple substitution, the detailed balance condition is satisfied as
\begin{equation}
\pi(\mathbf{x}) P(\mathbf{x},\mathbf{y}=\mathbf{y}_c)=\pi(\mathbf{y}) p(\mathbf{y}=\mathbf{y}_c,\mathbf{x}),
\end{equation}
completing the proof.
\end{proof}

\subsection{Convergence Analysis}

\begin{my2}
Given the invariant lattice Gaussian distribution $D_{\Lambda,\sigma,\mathbf{c}}$, the Markov chain induced by the independent MTMK sampling algorithm converges exponentially fast to the stationary distribution:
\end{my2}
\vspace{-1em}
\begin{equation}
\|P^t(\mathbf{x}, \cdot)-D_{\Lambda,\sigma,\mathbf{c}}(\cdot)\|_{TV}\leq \left(1-\delta_{\text{MTM}}\right)^t
\label{Uniform ergodicity of the multiple-try}
\end{equation}
\emph{with }
\begin{equation}\label{eq:delta-MTM}
\delta_{\text{MTM}}=\frac{k}{k-1+\frac{1}{\delta}}.
\end{equation}
The proof of Theorem 6 is provided in Appendix~\ref{prooft2}.

\begin{figure*}
\begin{center}
\includegraphics[width=4.4in]{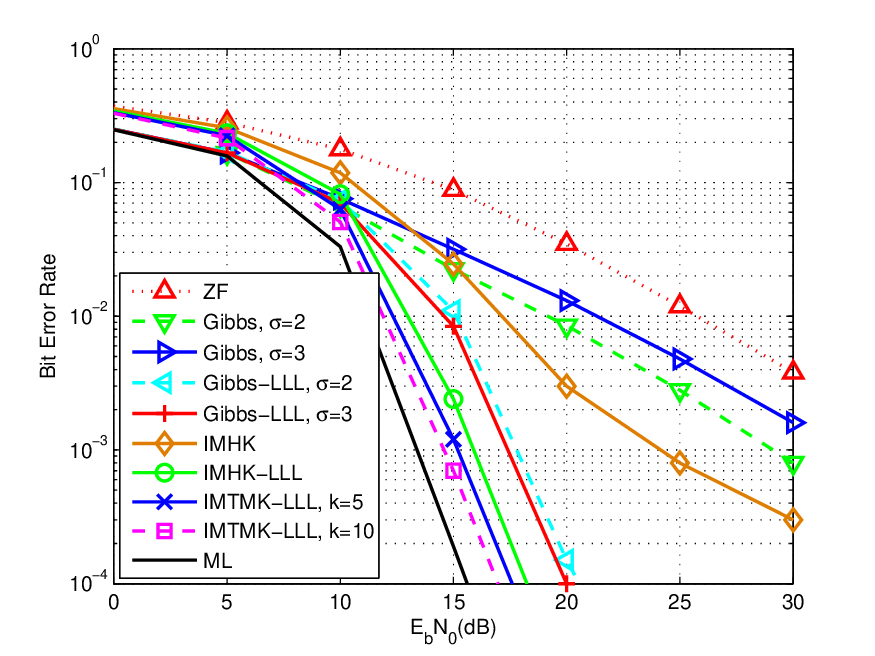}\\
\end{center}
\vspace{-1em}
\caption{Bit error rate versus average SNR per bit for the uncoded $8 \times 8$ MIMO system using 16-QAM  under 50 Markov moves.}
  \label{simulation 1}
\end{figure*}

From \eqref{eq:delta-MTM}, it can be observed that with the increase of the trial sample size $k$, the exponential decay coefficient $\delta_{\text{MTM}}=\frac{k}{k-1+\frac{1}{\delta}}$ will approach 1. In other words, with a sufficiently large $k$, sampling from the target distribution can be realized efficiently. More importantly, the generation of $k$ trial samples at each Markov move not only allows a fully parallel implementation, but also can be carried out in a preprocessing stage, which is beneficial in practice.

Now, given $\delta_{\text{MTM}}=\frac{k}{k-1+\frac{1}{\delta}}$, the mixing time of the underlying Markov chain can be estimated. Specifically, according to (\ref{mixing time}) and (\ref{Uniform ergodicity of the multiple-try}), we obtain
{\allowdisplaybreaks\begin{flalign}
t^{\text{MTM}}_{\text{mix}}(\epsilon)&=\frac{\text{ln}\hspace{.1em}\epsilon}{\text{ln}(1-\delta_{\text{MTM}})}\notag\\
&\overset{(g)}{<}\log\left(\frac{1}{\epsilon}\right)\cdot\left(\frac{1}{\delta_{\text{MTM}}}\right)\notag\\
&=\log\left(\frac{1}{\epsilon}\right)\cdot\left(\frac{k-1+\frac{1}{\delta}}{k}\right)\notag\\
&\approx\log\left(\frac{1}{\epsilon}\right)\cdot\left(\frac{1}{k\delta}\right),\ \ \epsilon < 1, \ \frac{1}{\delta}\gg k
\label{upperboundmixing1}
\end{flalign}}
where we again use the bound $\text{ln}(1-\alpha)<-\alpha$ for $0<\alpha<1$ in $(g)$. Clearly, the mixing time is proportional to $\frac{1}{k\delta}$, and becomes $O(1)$ if $k\delta \to 1$. Overall, compared with the mixing time given in (\ref{upperboundmixing}), the mixing time of the independent MTMK is significantly reduced by a factor of $k$. Since the independent MTMK inherits all the formulations of the independent MHK, we have
{\allowdisplaybreaks\begin{flalign}
C^{\text{MTM}}_{\mathrm{BDD}}&=\frac{t^{\text{MTM}}_{\text{mix}}(\epsilon)}{D_{\Lambda,\sigma,\mathbf{c}}(\mathbf{x})}\notag\\
&\lesssim\frac{1}{k}\cdot\frac{\log\left(\frac{1}{\epsilon}\right)\cdot\left(\frac{1}{\delta}\right)}{D_{\Lambda,\sigma,\mathbf{c}}(\widehat{\mathbf{x}})}\notag\\
&=\frac{1}{k}\cdot\log\left(\frac{1}{\epsilon}\right)\cdot C\notag\\
&=\frac{1}{k}\cdot\log\left(\frac{1}{\epsilon}\right)\cdot 1.0039^n \cdot e^{\frac{2\pi \cdot d^2(\Lambda, \mathbf{c})}{\min_i\|\widehat{\mathbf{b}}_i\|^2}}
\end{flalign}
for $\sigma= \min_{i}\|\widehat{\mathbf{b}}_i\|/(2\sqrt{\pi})$.

Following the afore-mentioned derivation, the decoding radius of the independent MTMK algorithm can be easily obtained as
\begin{equation}
R_{\text{MTM}}=\sqrt{\frac{1}{2\pi}\cdot\ln\frac{k C^{\text{MTM}}_{\mathrm{BDD}}}{a}}\cdot\min_{1\leq i\leq n}\|\widehat{\mathbf{b}}_i\|.
\label{d55}
\end{equation}

\begin{my5}
Although the independent MTMK algorithm is able to reduce the mixing time, its complexity in each move increases due to multiple calls of trial samples. Therefore, parallel implementation or preprocessing is highly desired to ease the complexity burden.
\end{my5}

Moreover, it is possible to have a varying $k$ at each Markov move, thereby resulting in an adaptive independent MTMK algorithm as
\begin{equation}
\|P^t(\mathbf{x}, \cdot)-D_{\Lambda,\sigma,\mathbf{c}}(\cdot)\|_{TV}\leq \prod_{i=1}^{t}(1-\delta^i_{\text{MTM}}),
\end{equation}
where $\delta^i_{\text{MTM}}=\frac{k_i}{k_i-1+\frac{1}{\delta}}$ and $k_i$ denotes the size of trial samples at each Markov move \cite{MartinoMTMIssue}.


\section{Experiments of MIMO Detection}
In this section, performance of the MCMC decoding algorithms is evaluated in MIMO detection. Specifically, we present simulation results for an $n\times n$ MIMO system with a square channel matrix. Here, the $i$th entry of the transmitted signal $\mathbf{x}$, denoted as $x_i$, is a modulation symbol taken independently from an $M$-QAM constellation $\mathcal{X}$
with Gray mapping. Meanwhile, we assume a flat fading environment, where the channel matrix
$\mathbf{H}$ contains uncorrelated complex Gaussian fading gains with unit
variance and remains constant over each frame duration. Let $E_b$ represents the average power per bit at the receiver, then the signal-to-noise ratio (SNR) $E_b/N_0=n/(\text{log}_2(M)\sigma_w^2)$ where $M$ is the modulation level and $\sigma_w^2$ is the noise power. Then, we can express the system model as
\begin{equation}
\mathbf{c}=\mathbf{H}\mathbf{x}+\mathbf{w}.
\label{eqn:System Model3}
\end{equation}
Typically, in the case of Gaussian noise $\mathbf{w}$ with zero mean and variance $\sigma_w^2$, it follows from (\ref{d21}) that
\begin{equation}
C\approx O(e^{2\pi n \sigma_w^2/\min_i\|\widehat{\mathbf{b}}_i\|^2})
\label{ddd}
\end{equation}
as $\|\mathbf{Bx}-\mathbf{c}\|^2\approx n \sigma_w^2$ by the law of large numbers. Therefore, the decoding complexity $C$ decrease with the SNR. Note that the noise variance $\sigma_w^2$ is different from the standard deviation $\sigma$ of the lattice Gaussian distribution\footnote{In \cite{HassibiMCMCnew}, the noise variance $\sigma_w^2$ is used as the sampling variance, but this would lead to a stalling problem at high SNRs \cite{MCMCKumarGibbsStalling}.}.



\begin{figure*}
\begin{center}
\includegraphics[width=4.4in]{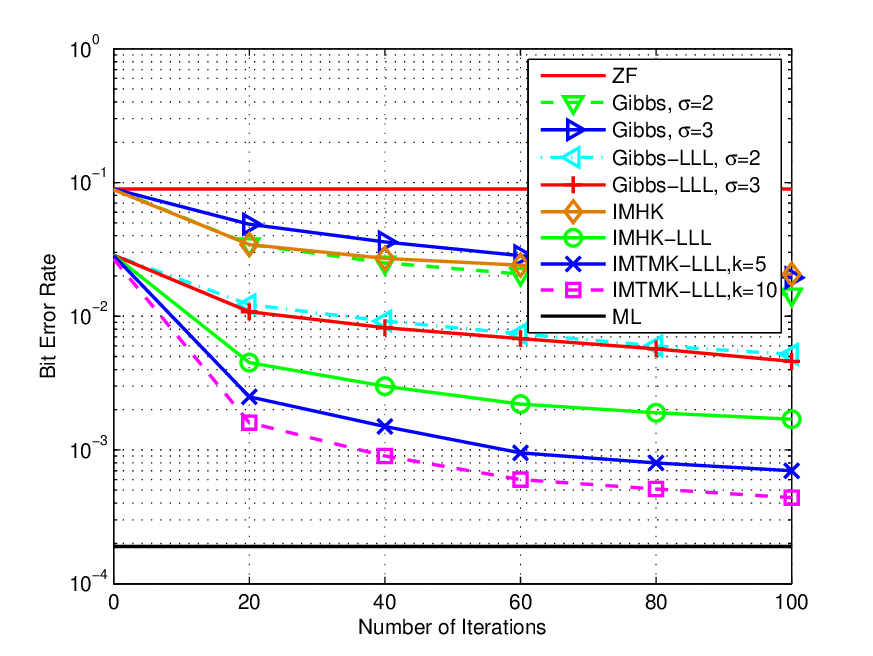}\\
\end{center}
\vspace{-1em}
\caption{Bit error rate versus the number of Markov moves for the uncoded $8 \times 8$ MIMO system using 16-QAM.}
  \label{simulation 2}
\end{figure*}

On the other hand, soft-output decoding for MIMO bit interleaver coded modulation (BICM) system is also possible using the samples generated by MCMC. Specifically, the sample candidates can be used to approximate the log-likelihood ratio (LLR), as in \cite{Hochwald}. For bit $b_i\in\{0,1\}$, the approximated LLR is computed as
\begin{equation}
L(b_i|\mathbf{c})=\text{log}\frac{\sum_{\mathbf{x}:b_{i}=1}\text{exp}\ (-\frac{1}{2\sigma^2}\parallel \mathbf{c}-\mathbf{H}\mathbf{x} \parallel^2)}{\sum_{\mathbf{x}:b_{i}=0}\text{exp}\ (-\frac{1}{2\sigma^2}\parallel \mathbf{c}-\mathbf{H}\mathbf{x} \parallel^2)},
\label{eqn:LLR}
\end{equation}
where $b_i$ is the $i$th information bit associated with sample $\mathbf{x}$. The notation $\mathbf{x}:b_{i}=\mu$ means the set of all vectors $\mathbf{x}$ for which $\mathbf{x}:b_{i}=\mu$.

Fig. \ref{simulation 1} shows the bit error rate (BER) of MCMC decoding in a $8\times8$ uncoded MIMO system with 16-QAM, where all the samples generated by MCMC algorithms are taken into account for decoding. This corresponds to a lattice decoding scenario with dimension $n = 16$. The performance of zero-forcing (ZF) and maximum-likelihood (ML) decoding are shown as benchmarks. For a fair comparison, sequential Gibbs sampling is applied here, which performs 1-dimensional conditional sampling of $x_i$ in a backward order\footnote{A forward update of $x_i$ in sequential Gibbs sampling is also possible.}, completing a full iteration \cite{HassibiMCMCnew}. This corresponds to one Markov move in the independent MHK and MTMK algorithms, which also update $n$ components of $\mathbf{x}$ in one iteration.

As expected, with $t=50$ Markov moves (i.e., iterations), independent MHK outperforms Gibbs sampling. As $\sigma$ has a vital impact on the sampling algorithms, Gibbs sampling is illustrated by tuning $\sigma$ with different values. Note that the detection performance may be affected due to the finite constellation.  Furthermore, as shown in (\ref{d5}), under the help of LLL reduction, the decoding radius of the independent MHK sampling is significantly strengthened by a larger size of $\min_{i}\|\widehat{\mathbf{b}}_i\|$, thereby leading to a much better decoding performance. As a comparison, LLL reduction is applied in Gibbs sampling as a preprocessing stage to yield the high quality initial starting point.
Additionally, compared to independent MHK, further decoding gain can be obtained by the independent MTMK algorithm, where cases with $k=5$ and $k=10$ trial samples are illustrated respectively.

On the other hand, in Fig. \ref{simulation 2}, the BERs of MCMC sampling detectors are evaluated against the number of Markov moves (i.e., iterations) in a $8\times8$ uncoded MIMO system with 16-QAM. The SNR is fixed as $E_b/N_0\hspace{-0.3em}=\hspace{-0.3em}15$ dB. Clearly, the performances of all the MCMC detectors improve with the number of Markov moves. Meanwhile, with the same number of Markov moves, a substantial performance gain is obtained by LLL reduction. By increasing number of trial samples, better decoding performance can be obtained by the independent MTMK algorithm due to a larger decoding radius shown in (\ref{d55}).


\section{Conclusions}
In this paper, the MCMC-based lattice Gaussian sampling was studied in full details. The spectral gap of the transition matrix of the independent MHK algorithm was derived and analyzed, which leads to a tractable exponential convergence rate of the Markov chain. A comparison with the extensions to Peikert's algorithm and rejection sampling illustrated the advantages of independent MHK. With the tractable mixing time, the decoding complexity of BDD using MCMC was derived  and a trade-off between the decoding radius and complexity was established. The potential of MCMC was further demonstrated in trapdoor sampling.
After that, by exploiting the potential of trial samples, the independent MTMK algorithm was proposed to enhance the convergence. It supports parallel implementation due to the independent proposal distribution, thus making independent MTMK algorithm promising in practice.



%

\section*{Acknowledgment}
The authors would like to thank Dr. Thomas Prest for helpful discussions.



\appendices

\section{Proof of Theorem 6}
\label{prooft2}

\begin{proof}
To begin with, let us take a careful look at the term $\text{min}\{\cdot,\cdot\}$ from (\ref{c2}). Here, for ease of presentation, we define
\begin{equation}
A=\hspace{-0.5em}\sum_{\mathbf{y}_{1:c-1}\in\mathbb{Z}^n}\sum_{\mathbf{y}_{c+1:k}\in\mathbb{Z}^n}\hspace{-0.5em}\left\{\frac{\prod_{j=1,j\neq c}^kq(\mathbf{y}_j)}{\omega(\mathbf{y}_c)+\sum_{j=1,j\neq c}^{k}\omega(\mathbf{y}_j)}\right\}
\end{equation}
and
\begin{equation}
B=\hspace{-0.5em}\sum_{\mathbf{y}_{1:c-1}\in\mathbb{Z}^n}\sum_{\mathbf{y}_{c+1:k}\in\mathbb{Z}^n}\hspace{-0.5em}\left\{\frac{\prod_{j=1,j\neq c}^kq(\mathbf{y}_j)}{\omega(\mathbf{x})+\sum_{j=1,j\neq c}^{k}\omega(\mathbf{y}_j)}\right\}.
\end{equation}

Meanwhile, because the $k$ trial samples from the proposal distribution $q(\cdot)$ are independent of each other, a set $\Xi$ is defined which contains the $k-1$ trial samples $\mathbf{y}_j$, $j\neq c$.

Then we can express $A$ and $B$ as
\begin{eqnarray}
A=\sum_{\Xi}Q(\Xi)\cdot{\frac{1}{\omega(\mathbf{y}_c)+\varpi(\Xi)}}=\sum_{\Xi}Q(\Xi)\cdot F_A(\Xi)
\end{eqnarray}
and
\begin{eqnarray}
B=\sum_{\Xi}Q(\Xi)\cdot{\frac{1}{\omega(\mathbf{x})+\varpi(\Xi)}}=\sum_{\Xi}Q(\Xi)\cdot F_B(\Xi).
\end{eqnarray}
Here, $Q(\Xi)=\prod_{j=1,j\neq c}^kq(\mathbf{y}_j)$ represents a probability distribution that takes all $q(\mathbf{y}_j)$, $j\neq c$ into account as a whole. On the other hand, $F_A(\Xi)$ and $F_B(\Xi)$ stand for the functions about $\Xi$, namely,
\begin{equation}
F_A(\Xi)={\frac{1}{\omega(\mathbf{y}_c)+\varpi(\Xi)}}
\end{equation}
and
\begin{equation}
F_B(\Xi)={\frac{1}{\omega(\mathbf{x})+\varpi(\Xi)}},
\end{equation}
where
\begin{equation}
\varpi(\Xi)=\sum_{j=1,j\neq c}^{k}\omega(\mathbf{y}_j).
\end{equation}
Now, let us focus on the term $A$, and we arrive at
{\allowdisplaybreaks\begin{flalign}
A&\hspace{.5em}=\hspace{.5em}\sum_{\Xi}Q(\Xi)\cdot F_A(\Xi)\notag\\
&\hspace{.5em}=\hspace{.5em}\mathbb{E}_{Q(\Xi)}[F_A(\Xi)]\notag\\
&\hspace{.5em}\overset{(h)}{\geq}\hspace{.5em}\frac{1}{\mathbb{E}_{Q(\Xi)}[\omega(\mathbf{y}_c)+\varpi(\Xi)]}\notag\\
&\hspace{.5em}=\hspace{.5em}\frac{1}{\omega(\mathbf{y}_c)+\mathbb{E}_{Q(\Xi)}[\varpi(\Xi)]}\notag\\
&\hspace{.5em}\overset{(i)}{=}\hspace{.5em}\frac{1}{k-1+\omega(\mathbf{y}_c)}.
\label{c3}
\end{flalign}}Here, $\mathbb{E}_{u(x)}[v(x)]$ represents the expectation of function $v(x)$ while $x$ is sampled from the distribution $u(x)$, $(h)$ comes from the \emph{Jensen's inequality} in the multi-variable case. Moreover, thanks to the $k-1$ independent samples from $q(\cdot)$, $(i)$ follows the derivations shown below,
{\allowdisplaybreaks\begin{flalign}
\mathbb{E}_{Q(\Xi)}[\varpi(\Xi)]&\hspace{.5em}=\hspace{.5em}(k-1)\cdot \mathbb{E}_{q(\mathbf{y}_j)}[\omega(\mathbf{y}_j)]\notag\\
&\hspace{.5em}=\hspace{.5em}(k-1)\cdot\sum_{\mathbf{y}_j\in\mathbb{Z}^n}q(\mathbf{y}_j)\cdot\omega(\mathbf{y}_j)\notag\\
&\hspace{.5em}=\hspace{.5em}(k-1)\cdot\sum_{\mathbf{y}_j\in\mathbb{Z}^n}\pi(\mathbf{y}_j)\notag\\
&\hspace{.5em}=\hspace{.5em}k-1.
\end{flalign}}Similar to $A$, we can rewrite $B$ as
\begin{equation}
B\geq \frac{1}{k-1+\omega(\mathbf{x})}.
\label{c4}
\end{equation}
Therefore, from (\ref{c3}) and (\ref{c4}), we get
{\allowdisplaybreaks\begin{flalign}
P(\mathbf{x}, \mathbf{y}=\mathbf{y}_c)&=k\cdot p(\mathbf{y}_c|\mathbf{x},c)\notag\\
&=k\cdot\pi(\mathbf{y}_c)\cdot\text{min}\{A,B\}\notag\\
&\geq\pi(\mathbf{y}_c)\cdot\text{min}\left\{\frac{k}{k-1+\omega(\mathbf{y}_c)},\frac{k}{k-1+\omega(\mathbf{x})}\right\}\notag\\
&\geq\pi(\mathbf{y}_c)\cdot\frac{k}{k-1+\omega_{\text{max}}(\mathbf{x})}\notag\\
&\geq\pi(\mathbf{y}_c)\cdot\frac{k}{k-1+\frac{1}{\delta}}\notag\\
&=\delta_{\text{MTM}}\cdot\pi(\mathbf{y}_c),
\label{c5}
\end{flalign}}where $\delta_{\text{MTM}}=k/(k-1+\frac{1}{\delta})$ and
{\allowdisplaybreaks\begin{flalign}
\omega_{\text{max}}(\mathbf{x})&\triangleq\text{sup}\ \omega(\mathbf{x})=\text{sup}\ \frac{\pi(\mathbf{x})}{q(\mathbf{x})} \notag\\
&\leq\frac{1}{\delta}
\end{flalign}}for $\mathbf{x}\in\mathbb{Z}^n$ from (\ref{xxxx}) in Lemma 1. From (\ref{c5}), it is straightforward to see that all the Markov transitions have a component of size $\delta_{\text{MTM}}$ in common. Then, uniform ergodicity can be easily demonstrated through spectral gap or coupling technique, which is omitted here for simplicity.

\end{proof}

\bibliographystyle{IEEEtran}
\bibliography{IEEEabrv,reference1}

\end{document}